\documentclass[a4paper,fleqn,usenatbib]{mnras}

\usepackage{newtxtext,newtxmath}

\usepackage[T1]{fontenc}
\usepackage{ae,aecompl}

\usepackage{graphicx}	
\usepackage{amsmath}	
\newcommand{\xten}[1]{\mbox{$\times 10^{#1}$}}

\newcommand{\ltappeq}{\raisebox{-0.6ex}{$\,\stackrel
{\raisebox{-.2ex}{$\textstyle <$}}{\sim}\,$}}
\newcommand{\gtappeq}{\raisebox{-0.6ex}{$\,\stackrel
{\raisebox{-.2ex}{$\textstyle >$}}{\sim}\,$}}

\title[Quasi-equilibrium chemical evolution]
{Quasi-equilibrium chemical evolution in starless cores}

\author[J.M.C. Rawlings, E. Keto and P. Caselli]
{J.M.C.~Rawlings,$^{1}$\thanks{E-mail: jcr@star.ucl.ac.uk}
E.~Keto,$^2$ and P.~Caselli$^3$\\
\\
$^{1}$Department of Physics and Astronomy, University College London,
Gower Street, London, WC1E 6BT, UK\\
$^2$Harvard-Smithsonian Center for Astrophysics, 160 Garden St., Cambridge,
MA 02420, USA\\
$^3$Max-Planck-Institut f\"{u}r extraterrestrische Physik, P.O. Box 1312, 
D-85741 Garching, Germany}

\date{Accepted 2024 April 21. Received 2024 April 08; in original form 
2023 November 17}

\pubyear{2023}

\begin{document}
\label{firstpage}
\pagerange{\pageref{firstpage}--\pageref{lastpage}}
\maketitle

\begin{abstract}
The chemistry of H$_2$O, CO and other small molecular species in 
an isolated pre-stellar core, L1544, has been assessed in 
the context of a comprehensive gas-grain chemical model, coupled to 
an empirically constrained physical/dynamical model.
Our main findings are (i) 
that the chemical network remains in near equilibrium as the 
core evolves towards star formation and the molecular abundances change 
in response to the evolving physical conditions. 
The gas-phase abundances at any time can be calculated accurately
with equilibrium chemistry, and the concept of chemical clocks is
meaningless in molecular clouds with similar conditions and dynamical 
time scales, and  
(ii) A comparison of the results of complex and simple chemical networks
indicates that the abundances of the dominant oxygen and carbon species, 
H$_2$O, CO, C, and C$^+$ are reasonably approximated by simple networks. 
In chemical equilibrium, the time-dependent differential terms vanish and 
a simple network reduces to a few algebraic equations. This allows rapid 
calculation of the abundances most responsible for spectral line radiative 
cooling in molecular clouds with long dynamical time scales. 
The dust ice mantles are highly structured and the ice 
layers retain a memory of the gas-phase abundances at the time of their
deposition.
A complex (gas-phase and gas-grain) chemical structure therefore exists, 
with cosmic-ray induced processes dominating in the inner regions. 
The inferred H$_2$O abundance profiles for L1544 require that the outer parts 
of the core and also any medium exterior to the core are 
essentially transparent to the interstellar radiation field.
\end{abstract}



\section{Introduction}
\label{sec:intro}

In this paper, we seek to improve our modelling and understanding of the chemistry 
of quiescent molecular gas with our astrochemistry code, {\sc STARCHEM}
\citep[][and described more fully in Appendix~\ref{sec:starchem}]{RW21,Raw22}. 
The {\sc STARCHEM code} is designed to integrate chemical evolution along Lagrangian streamlines of fluid flow. In this paper, we apply {\sc STARCHEM} in conjunction with 
a previously developed dynamical model for the gravitationally contracting molecular 
cloud L1544 \citep[][hereafter KC10 and KCR15, respectively]{KC10,KCR15}. 
Our goals in this study are twofold.

Firstly, we test the hypothesis that the chemical evolution can be treated as being
in quasi-equilibrium, so that the gas-phase abundance distributions are simply 
determined by the instantaneous physical conditions. 
As part of this analysis, we pay particular attention to the chemical stratification 
within the dust grain ice mantles, and so differentiate between the gas-phase 
chemistry and that of the ices, which record the chemical evolution of the core.

Secondly, we assess the accuracy and applicability of a radically simplified 
chemical model developed in a series of earlier papers; KC10 and \citet{KC08,KRC14} 
(hereafter KC08 and KRC14, respectively). 
This model is one of several proposals in the literature for reduced chemical 
networks that are also useful in rapid computation of molecular gas temperatures 
\citep{NL97,GC12}.
They directly compute the abundances of the important coolants such as CO, C$^+$, 
C and O. Abundances of other molecular species can be estimated from those 
calculated to develop a general cooling function \citep{Gold01,KF05}. 
Because of their computational speed, these reduced 
chemical networks are useful in numerical, hydrodynamic simulations of the 
ISM which are already computationally intensive without the additional burden 
of a time-dependent complex chemistry with many reactions. An earlier 
version of our reduced chemical network has been incorporated into one 
such code \citep{BK15}.

We choose L1544 as an example because we have an observationally verified model 
for the dynamics of this cloud that provides specific combinations of gas and 
dust temperatures,  gas density, and UV radiation for our simulations of the astrochemistry. In addition, the different radial zones in the cloud, outer, 
central, and transition, provide examples where the chemistry operates in very 
different ways. Whilst L1544 is a particular cloud, its chemistry is 
representative of quiescent molecular gas in the ISM meaning gas that is not 
strongly disturbed by shocks, bipolar outflows, intense radiation from 
nearby stars, or processes in accretion disks.

In the following section we discuss the structure and dynamics of the L1544 
pre-stellar core. Our dynamical/chemical model is described in section \ref{sec:chem}. 
In section \ref{sec:results} we present the results discussing (a) the chemical 
structure of the  cloud (b) the structure of the ice mantles (c) the chemical 
time-dependence, and (d) the validity of reduced chemical networks. 
Our conclusions are summarized in section section \ref{sec:conc}. A proposed 
revised form of our reduced chemical network is given in 
Appendix~\ref{sec:CO_lite} and the main results of STARCHEM applicable to 
starless and pre-stellar cores are described in Appendix~\ref{sec:starchem}.

\section{The physical structure and dynamics of L1544}
\label{sec:struct}

The pre-stellar core L1544 is located in the Taurus Molecular Cloud, at a 
distance of $\sim$170pc \citep{Galli19}. Infrared observations show no evidence 
of an embedded protostar; however, split, self-absorbed spectral line profiles 
of several molecules that are tracers of dense gas indicate that the central 
region is in the early stages of gravitational collapse toward star formation. 
On a larger scale, molecular line observations show patchy emission around the 
core suggesting that the core is isolated within the Taurus cloud with no 
evidence for surrounding filamentary structure or other cores overlapping 
along the line of sight. The overall morphology is roughly elliptical with 
an aspect ratio of 2:1; however contours of constant density become more 
spherical with increasing density suggesting that the dynamics can be 
understood in the context of a spherical model \citep[e.g.][]{Cas22}. 
With these properties, L1544 has become a prototypical example of a small 
molecular cloud that is just beginning gravitational collapse towards the 
formation of a solar-mass scale star. 

The chemistry within L1544 is both relatively simple and interestingly 
complex. The absence of a protostar implies the absence of the chemistry 
associated with shocks, a bipolar outflow, and an accretion disk. The chemistry 
is consistent with that of quiescent molecular gas. With no embedded protostar 
or nearby bright stars, the core is illuminated only by the external radiation 
field from Galactic starlight. This sets up a radial structure within the 
core of varied chemistry. 
Photochemistry dominates the outer region, but the inner core is cold 
\citep[5-10K,][]{Crap07} and dark, so that the chemistry is mostly 
restricted to grain surfaces and the 
gas-grain interactions of freeze-out and desorption. L1544 is in a family of 
starless cores in which CO has been highly depleted \citep[e.g.][]{Cas99,BT07}.
Indeed, recent high resolution observations have shown almost completele 
depletion in the central regions of the core 
\citep{Cas22}.

Along with its advantages for theoretical study, L1544 is observationally 
favorable. At 170 pc distant, the core is extended over arc minute scales 
and well matched to the angular resolution of single dish radio telescopes 
at millimeter and submillimeter frequencies. A number of spectral lines has 
been observed with varying optical depths from molecules with different 
critical densities for collisional de-excitation which provide information 
on both the inner and outer regions of the core. While the 
dynamics and the chemistry can be studied independently, it is not possible 
to understand the observations without consideration of both. The most 
obvious effect is that the molecular abundances and gas velocity are both 
functions of radius resulting in differences in the self-absorbed line 
profiles of different molecules. Once understood, this effect provides precise 
estimates of the emission and absorption regions for each molecule. 

In a series of earlier papers we developed a model for the chemistry and 
dynamics of L1544 that has been successful in both matching and predicting 
the spectral line shapes of several molecules. Originally based on 
observations of CO and N$_2$H$^+$, the model was able without modification to 
predict the blue-shifted self-absorbed emission of H$_2$O and match the 
complex pattern of the self-absorbed hyperfine lines of NH$_3$ 
\citep[KRC14,][]{Cas17}.

The physical model of L1544 is described in KC10, as resulting from 
the quasi-static contraction of a thermally supercritical 10M$_\odot$ cloud, 
of radius 0.75pc that is initially configured as a Bonnor-Ebert sphere, in 
unstable dynamical equilibrium, with a central density of 
$2\times 10^4$cm$^{-3}$. 
After contraction to a central density of $2\times 10^7$cm$^{-3}$, the 
structure of the model describes the observed structure of L1544. Subsequent 
observations suggest minor differences \citep[e.g.][]{Bizz14}.
\citet{Chac19} proposed that the `flat central region' is more extended and 
has a lower density ($\sim 2\times 10^6$cm$^{-3}$) than that of KC10 - based 
on 1.1mm Large Millimetre Telescope and 3.3mm Green Bank Observatory data.
Based on 1.3mm ALMA data, \citet{Cas19} found that it is somewhat smaller, 
elongated, but consistent with a central density of $\sim 1\times 10^7$cm$^{-3}$.
Observationally, the density in the center of the core cannot be determined
directly, only the column density through the centre. The suggested 
differences are consistent with observational uncertainty.

\begin{figure*}
\includegraphics[scale=0.7]{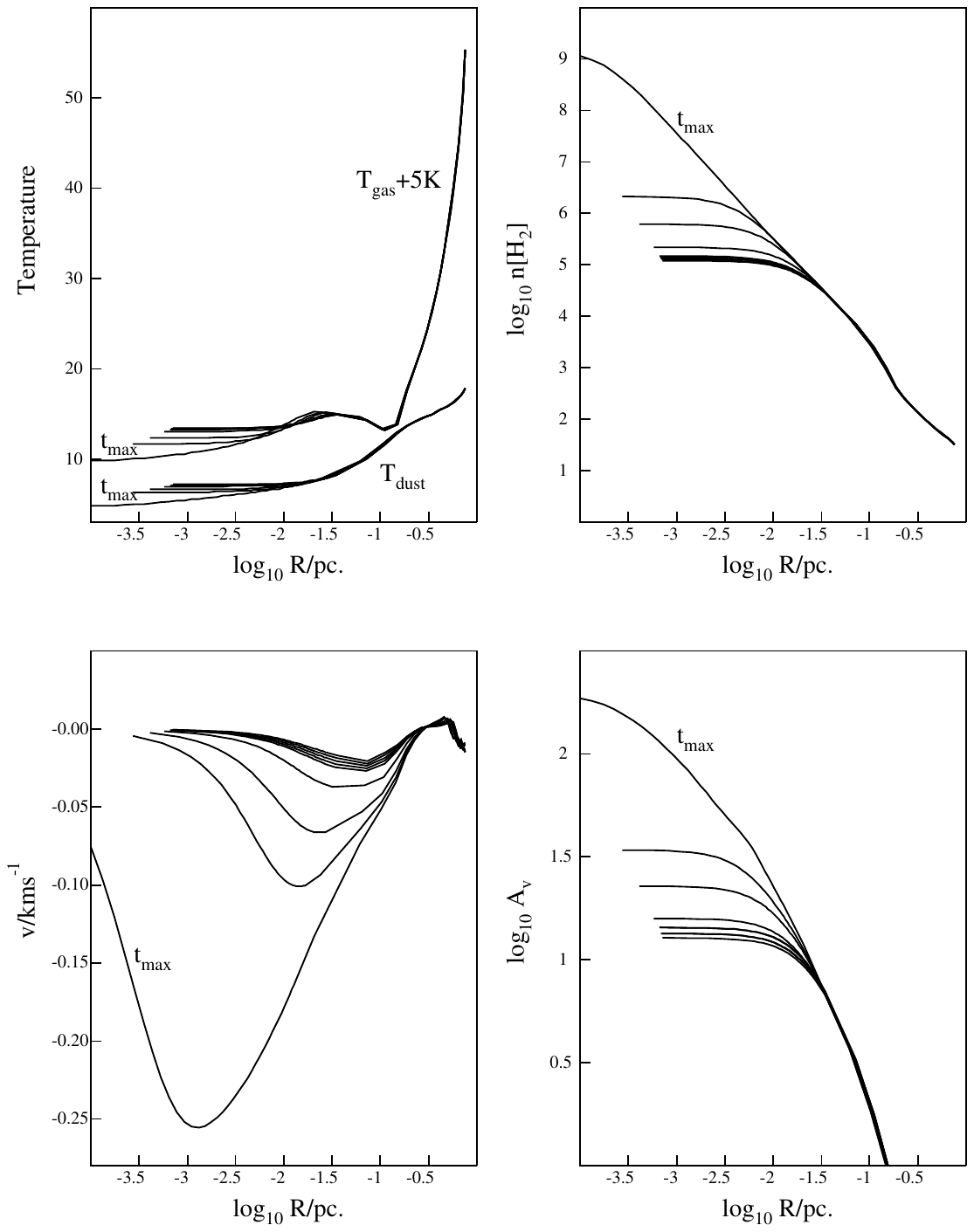} 
\caption{The physical model of L1544, as described in \citet{KC10}. 
The figure shows the radial profiles of the gas and dust temperatures 
($T_{\rm gas}$ and $T_{\rm dust}$), the H$_2$ density, the radial velocity and the 
direction-averaged extinction, at several times between $t=t_{\rm min}$ and 
$t=t_{\rm max}$. The curves for $t=t_{\rm max}$ are labelled to indicate
the direction of time. The curves for $T_{\rm gas}$ have been elevated by 5K to 
avoid confusion with those for $T_{\rm dust}$. 
$T_{\rm gas}$ and $T_{\rm dust}$ are well-coupled in the central regions.}
\label{fig:physics}
\end{figure*}

Figure~\ref{fig:physics} shows the temperature (gas and dust), density, 
velocity and extinction profiles; $T_{\rm gas}(r)$, $T_{\rm dust}(r)$, 
$n(r)$, v$(r)$ and $A_{\rm v}(r)$, at several times between 0.30Myr 
($t_{\rm min}$) and 0.61Myr ($t_{\rm max}$) since the start of the 
contraction from the initial configuration. The profiles for 
$t=t_{\rm max}$ (which are labelled and indicate the direction of time) 
correspond to those published as the best fits in 
\citet{KRC14} (hereafter KRC14) and subsequently used in other studies.
The A$_{\rm v}(r,t)$ profiles give the, astrochemically 
relevant, mean direction-averaged extinction at any particular point 
and time within the spherical model of the core.
\footnote{Since starless cores have no internal sources of 
radiation, the radiation field at any point in the core is the radiation 
field exterior to the core diminished by the mean Av averaged over all 
directions to the boundary of the core. The mathematical definition of 
the average can be found in Appendix 1 of \citet{ZWG01}.}
This is not the same as the, more observationally relevant, line of
sight extinction - which is what is depicted in, for example, figure 3 of KRC14.

From Figure~\ref{fig:physics} we can see that the density profiles progress 
towards $n\propto r^{-2}$
as the central region of constant density defined by the radius where 
the crossing time equals the free-fall time decreases in size.
For L1544, the profiles for $t=t_{\rm max}$ (and, in particular, the density
and velocity profiles, and the value of the central density 
$\sim 10^{7}$cm$^{-3}$) are empirically constrained by the observed 
H$_2$O 576GHz ($1_{10}-1_{11}$) line profile. This density corresponds to a 
{\em specific time} ($t=t_{\rm max}\sim$0.61Myr) in the dynamical evolution, 
as can be seen from the time-dependence of the velocity profiles. At this time 
the dynamical evolution of the cloud is beginning to accelerate, with 
infall velocities just becoming supersonic in the inner region 
(r$\ltappeq$0.01pc, $A_{\rm v}\gtappeq 23$). 
It is also evident that the gas and dust temperatures are reasonably 
well-coupled within $\sim$0.15pc ($A_{\rm v}\gtappeq 0.96$), and are 
very low throughout most of the cloud. 
This has significant implications for the chemistry which we discuss in
section~\ref{sec:results}.


Although the core evolves very slowly, the physical model nonetheless 
implies an important time-dependence, so that the best fit to the 
observations and the implied conditions representing the present status of 
a core are defined by a specific dynamical age. 
%
In general, there are essentially two possibilities: 
(i) the timescale for chemical evolution ($\tau_{\rm chem}$) is less than 
that for the dynamical evolution ($\tau_{\rm dyn}$), 
in which case quasi-equilibrium chemical evolution will hold - with the 
abundances responding to the changes in the instantaneous physical conditions, or 
(ii) $\tau_{\rm chem}>\tau_{\rm dyn}$, in which case the chemistry will 
depend on both the dynamical evolution and the initial conditions.
In either case, a static chemistry is not a realistic representation.

%

The fastest timescale in which 
a Bonnor-Ebert configuration in hydrostatic equilibrium can be assembled 
via sub-sonic quasi-static contraction is of the order of the free-fall 
collapse timescale:
\[ t_{\rm ff} = \sqrt{\frac{3\pi}{32G\rho}} = 3.1\times 
10^5\left(\frac{10^4{\rm cm}^{-3}}{n_{\rm H_2}}\right)^{1/2}~{\rm years} \]
where $n_{\rm H_2}$ is the (assumed molecular) gas density and we have assumed 
that  He/H =0.1.
Furthermore, as the core {\em must} be in near-hydrostatic equilibrium
(KCR15) its age must be greater than the sound-crossing time 
($t_{\rm cross}$). 
Even if we only consider the inner 0.1pc then, for a core composed of 
molecular gas, with mean temperature $\sim$10K; $t_{\rm cross}>1$Myr. 
Shorter timescales than these could only be obtained as a result of 
supra-gravitational, supersonic contraction, 
which would occur only if the core were out of equilibrium everywhere. 
As shown in KCR15 an increase of the density by $\sim 10$\% over its 
equilibrium values along the profile would result in global cloud collapse 
and a velocity structure that is incompatible with the subsonic spectral 
line profiles of all observed molecules other than H$_2$O.
Moreover, as can be inferred from Fig.1, if the dynamical age were only 
slightly larger than this, then much higher central densities, together with 
larger inflow velocities throughout the envelope would result.

It has taken $t_{\rm form}\sim$0.6Myr to 
construct the existing configuration from an initial state in dynamical 
equilibrium that was marginally unstable 
and (from the implied velocity profile) that it is now evolving on a near 
gravitational free-fall time-scale of $<0.3$Myr.
Thus $t_{\rm form}$ is a fixed parameter. Any model of the 
chemistry that identifies a nominal `chemical age' that is independent of 
the physical evolution would be quite inconsistent with this, or any 
other realistic, dynamical model. The inapplicability of hydrodynamically
static chemical evolution was also noted by \citep{SCRS22} in their 
combined chemical-hydrodynamical study of the velocity structure in L1544. 

We should also consider the physical and chemical 
evolution of the cloud {\em prior} to the formation of the initial 
Bonnor-Ebert sphere, since parts of this cloud will already have relatively 
high extinctions and low temperatures, implying that they must have previously 
evolved through a PDR interface.
Moreover, in the circumstances where the chemical evolution is slow 
($\tau_{\rm chem}>\tau_{\rm dyn}$), then a larger effective chemical age could
be obtained if it is assumed that the core maintains hydrostatic equilibrium 
for some time {\em before} quasi-static
contraction starts, so that the chemical evolution occurs at relatively 
low densities (with correspondingly longer chemical timescales).

\section{The chemical model of starless cores}
\label{sec:chem}

To understand the chemical structure and evolution of isolated starless
cores we utilise
the {\sc STARCHEM} dynamical/chemical model as described in Appendix B.

When modelling the astrochemistry of pre- and proto-stellar cores, 
the usual approach is to study the time-dependent
chemical evolution, but freezing the physical conditions at $t=t_{\rm max}$, 
with the inclusion of some chemical pre-evolution at constant density.
As is clear from the discussion in the previous section, this is quite arbitrary 
and incompatible with our understanding of how the source has evolved.
%
%
This has been noted in previous studies (KC10, KRC14) in the context 
of the time-dependences of the abundance profiles for CO and H$_2$O in L1544, 
this important fact has not been considered in most subsequent modelling efforts.

Moreover, even if we make the assumption that chemical quasi-equilibrium holds
(i.e. in response to the changing physical conditions), then the abundances
will be determined by the equilibrium balance between the gas-phase chemistry 
and the chemistry in the surface layers of the ice mantles on dust grains.
In conditions where significant freeze-out is occurring then the sub-surface 
layers of the ices - which are effectively chemically inert - will be determined 
by the entire evolutionary history of the cloud/core.  

We go some way towards addressing these inconsistencies and 
assumptions by following the combined dynamical and chemical evolution between 
$t_{\rm min}$ and $t_{\max}$, although we also make the simplification
that the core `instantly assembles' in the configuration specified at
$t=t_{\rm min}$.
Indeed, even if the initial configuration is taken to be a sphere 
of uniform density, that sphere will itself have position-dependent 
temperature and extinction profiles that will result in significant, 
and time-dependent, chemical structure. A fully self-consistent approach
would involve a knowledge of the density, temperature and extinction 
profiles from the very earliest stages.

%
{\sc STARCHEM} considers a full multi-point description of the time-dependent
chemistry and is configured to co-integrate the chemical and 
physical evolution in Lagrangian streamlines of 100 grid points. These  
are (initially) equally spaced between 1000au ($r_{\rm min}$) and 0.3pc 
($r_{\rm max}$) with an assumed spherical (1D) symmetry
(corresponding to $0.33\ltappeq A_{\rm v}\ltappeq 40$). 
The value of $r_{\rm min}$ is chosen so that $r>0$ at the end of the 
simulation, whilst $r_{\rm max}$ corresponds to the physical size of 
the core.

The physical conditions are defined according to the hydrodynmic model of 
KC10 and consist of the density $n(r,t)$, inflow velocity $v(r,t)$, 
gas temperature $T_{\rm gas}(r,t)$, dust temperature $T_{\rm dust}(r,t)$ 
and the mean visual extinction to the surface of the cloud $A_{\rm v}(r,t)$. 
The look-up tables give data as functions of positions and time for 1000 radial 
points (defined in the range $6.2au < r < 0.75pc$, corresponding to
$0.07\ltappeq A_{\rm v}\ltappeq 200$) at 62 times (in the range 
$3.0\xten{5}<t<6.1\xten{5}$ years). 

For times prior to $t_{\rm min}=3\xten{5}$ years the core evolves very slowly 
and, we take it to be dynamically static so that
we set the parameters to their boundary values, 
i.e. $n(r)=n(r,t_{\rm min}), $v$(r)=0, T(r)=T(r,t_{\rm min})$, and 
$A_{\rm v}(r)=A_{\rm v}(r,t_{\rm min})$.
In effect, there are two dynamical phases: (i) hydrostatic equilibrium from 
$t=0 - t_{\rm min}$, then (ii) quasi-static contraction from 
$t=t_{\rm min}$ to $t=t_{\rm max}$. This is a satisfactory approach, so long as 
the chemical timescale is shorter than the dynamical timescale.
As this is debatable, we have included the possibility of an additional 
dynamical `pause' ($t_{\rm pause}$) at the beginning of the simulation to 
test the assumption.

Gas-phase reactions, freeze-out, various continuous desorption mechanisms 
and surface chemical reactions are all important processes in starless cores.
We include a comprehensive gas-phase chemistry and full descriptions
of the desorption mechanisms, coupled to a fully time and position dependent
analysis of the compositional structure of the ices.
Photochemistry plays an important role in the chemical evolution
of starless cores, both through gas-phase photolysis and the photodesorption
mechanisms, that depend on the spatial
variations of the radiation field.
The outer parts of the core ($r\gtappeq 0.07$pc, $A_{\rm v}\ltappeq 3.1$) 
are effectively in a photon dominated region (PDR), where the mutual shielding 
of the H$_2$, C and CO photodissociation and photoionization continua, together 
with dust absorption 
attenuates the radiation field as it penetrates the cloud. 
Allowances for these various attenuation effects are included in the model.
The model incorporates a simplified form of the surface chemistry, explained 
and justified in the appendix.

We follow the chemistry of some 81 gas-phase and 17 solid-state 
chemical species (of limited molecular complexity, containing no more than 
six atoms) composed of the elements H, He, C, N, O, S and Na (which we use as a 
`representative' low ionization potential metal). 
The elemental abundances relative to hydrogen are given in 
Table~\ref{tab:param} and are typical values for low mass star-forming regions
\citep{vD21}. 

The interaction between gas-phase atoms and molecules and the surface of
dust grains is of crucial importance to our understanding of the chemical
structure of molecular clouds and star-forming regions.
Notwithstanding the discussion above concerning the 
physics/dynamics, this is by far the most important source of
ambiguity and possible misinterpretation of the observational data.
In general, we need to consider the processes of freeze-out, or depletion
of a species from the gas-phase, surface chemistry, and the back-process of 
desorption, or sublimation, of the species from the (surface) of dust grains, 
back into to the gas-phase.
With the exception of H$_2$O, the desorption properties are not well defined for
the dominant ice species, CO, NH$_3$, CO$_2$ and CH$_4$, as well as N$_2$, 
O$_2$ and the monomers, C, N and O, from which the hydrogenated species form.

We concentrate on CO and H$_2$O, whose abundance profiles 
are reasonably well-constrained. Although this could be extended to the related 
species CO$_2$, which is a major ice component, an in-depth analysis is
not justified due to the lack of empirical constraints for this species 
in L1544.

Desorption of ices by the cosmic ray heating of grains dominates the 
desorption in the darkest parts of the core. However, the rates are uncertain.
Recent studies \citep[e.g.][]{SSC21,Raw22} have shown that when realistic dust 
grain cooling profiles are considered the rates are typically an order
of magnitude faster than those determined in previous calculations \citet{HH93}.
The rates may be significantly larger if the effects of sporadic desorption,
ice species co-desorption and/or whole mantle evaporation from small grains 
are included.  
In this study we use the revised rates for a standard interstellar dust grain 
size distribution \citep[][hereafter MRN]{MRN77} given in Table 3 of \citet{Raw22}.
KC10 found that it was necessary to use a CO desorption rate of 
$k_{\rm crdes}^{\rm CO}\sim 3\times 10^{-13}$s$^{-1}$ to explain the 
high CO abundance that is observationally inferred for the central regions.
We use a value of $k_{\rm crdes}^{\rm CO}=5.9\times 10^{-13}$s$^{-1}$, 
appropriate for CO molecules with a binding energy of 1100K, and a mean 
adsorbate binding energy of 1200K, \citep[][Table 4]{Raw22}. This 
value is very close to that determined by \citet{HC06} in their detailed 
random walk Monte Carlo model of the desorption mechanism
\citep[see also][]{Sil21}.

\subsection{Model parameters}
\label{subsec:freepara}

For our standard model we match the conditions in L1544, as the prototypical
example of an isolated pre-stellar core. 
The values that we use for the various parameters are given in 
Table~\ref{tab:param}.
Being marginally thermally supported with no internal energy source, the 
physical structure of L1544 is actually relatively simple 
so that these are fairly well-constrained and treated as fixed 
in our physical-chemical model. 
The values of some of the parameters (identified in Table~\ref{tab:param})
are defined by the chemical abundance profiles and we have considered the 
sensitivity of our results to these (see section~\ref{sec:free}).
%

\begin{table*}
\caption{Parameters in the model and values adopted. 
Those in the upper half of the table are fixed, or semi-constrained, 
whilst those in the lower half are the key free parameters and the values
given are for the best-fit model. See descriptions in the text.}
\begin{tabular}{|l|l|}
\hline
 Carbon abundance ($C/H$) & $1.5\times 10^{-4}$ \\
 Nitrogen abundance ($N/H$) & $7.4\times 10^{-5}$ \\
 Oxygen abundance ($O/H$) & $2.5\times 10^{-4}$ \\
 Sulfur abundance ($S/H$) & $1.0\times 10^{-7}$ \\
 Initial radial grid extent & 1000au - 0.3 pc \\ 
 Final radial grid extent & $\sim$100au - 0.3 pc \\
 Grain albedo & 0.5 \\
 Sticking coefficient ($S_{\rm i}$) & 1.0 (all species)\\
 Interstellar radiation field photon flux ($F_0$) & 1.0$\xten{8}$ photons cm$^{-2}$ s$^{-1}$ \\
 Cosmic ray induced photon flux ($F_{\rm cr}$) & 4875.0 photons cm$^{-2}$ s$^{-1}$ \\
 Binding site surface density ($N_{\rm s}$) & 1.0$\xten{15}$ cm$^{-2}$ \\
 Mean ice monolayer thickness ($\Delta r_{\rm layer}$) & 3.7\AA \\
\hline
 Hydrodynamic pause duration ($t_{\rm pause}$) & 0 years \\
 ISRF/Draine ($G_0$) & 1.0 \\
 External extinction ($A_{\rm v,ext.}$) & 0.0 magnitudes \\
 Cosmic ray ionization rate for H$_2$ ($\zeta_{\rm cr}$) & 1.3$\xten{-17}$ s$^{-1}$ \\
 Dust surface area per H nucleon ($\sigma_{\rm H}$) & 6.0$\xten{-21}$ cm$^2$ \\
 RMS grain radius ($a_{\rm rms}$) & 0.01$\mu$m \\
 Photodesorption yield ($Y_{\rm pd}$) & 1.0$\xten{-3}$ (see text) \\
 Yield for H$_2$-formation driven desorption ($Y_{\rm H_2}$) & 0.0 \\
 Fraction of sticking O/O$^+$/OH that reacts on grain surfaces ($F_{\rm reac}$) & 1.0 \\
 Fraction of surface-formed OH that chemically desorbs ($D_{\rm OH}$) & 0.25-0.5 (see text) \\
 Fraction of OH(s) that reacts with CO(s) to form CO$_2$(s) ($F_{\rm CO_2}$) & 0.1 \\
 Fraction of surface-formed H$_2$O that chemically desorbs ($D_{\rm H_2O}$) & 0.3-0.8 (see text) \\
 Fraction of O$_2$(s) that hydrogenates to H$_2$O(s) ($F_{\rm O_2}$) & 1.0 \\
\hline
\end{tabular}
\label{tab:param}
\end{table*}

Since the various surface processes depend on the grain surface area, 
the rms grain radius ($a_{\rm rms}$) for the grain size distribution
merits discussion. 
The gas-grain interaction and the microphysics of the evolution of
the ice mantles (as described in Appendix~\ref{sec:starchem}) should be 
followed for grains of different size. Here, as in other studies, we instead 
represent the distribution by mean values of the grain radius, 
cross-section etc.
Conventionally, astrochemical models use a `classical' value of
$a=0.1\mu$m. However, the empirically constrained MRN interstellar grain
size distribution is characterised by a power-law:
$n(a)\propto a^{-3/5}$ with lower and upper limits of
$a_{\rm min}\sim 0.005\mu$m and $a_{\rm max}\sim 1.0\mu$m.
The surface area in this distribution is heavily weighted towards the small 
grain population and the resulting $a_{\rm rms}\sim 0.011\mu$m. 
These small grains also dominate the UV opacity that regulates the 
photoprocesses.

A larger value of $a_{\rm rms}$ would be more appropriate if the effects of 
dust grain coagulation and/or ice mantle accretion are significant, as may be
appropriate in dense, cold gas \citep{Sil20}.
However, if this were the case, then there would be several other
implications to consider. These include
(a) the effects on the optical properties of the dust, specifically its
UV opacity - which affects the rates of the $A_{\rm v}$-dependent photoprocesses, 
(b) the effect on the grain surface area per unit volume,
(c) the electrostatic enhancement factor for the freeze-out of ions, and
(d) the relative rate at which dust grains grow, due to the accumulation 
of ice mantles (Rawlings et al., {\em in preparation}). 
Nonetheless, for the purpose of this study
and for the sake of consistency and comparison with previous models, 
we adopt the values of $\overline{a}=a_{\rm rms}=0.01\mu$m as given in 
Table~\ref{tab:param}, and defer a full discussion of these issues to a 
future study.

The value of the dust surface area per hydrogen nucleon ($\sigma_{\rm H}$) 
depends on a number of factors, including the assumed value of the dust to gas 
ratio (by mass, $d_{\rm m}$), $a_{\rm min}$, and the mean density of the dust 
grains ($\rho_{\rm g}$). 
With these uncertainties, $\sigma_{\rm H}$ for the MRN distribution lies
in the range $3-8\times 10^{-21}$cm$^{2}$.
Assuming that $d_{\rm m}\sim 0.1$ (by mass) and 
$\rho_{\rm g}\sim 3.0$g\,cm$^{-3}$ then 
$\sigma_{\rm H}\sim 6.6\times 10^{-21}$cm$^{2}$.
$\sigma_{\rm H}$ is an important parameter in our model and 
we adopt a standard value of $6.0\times 10^{-21}$cm$^{2}$ 
(which is the same as used in KRC14).

The photophysics can be moderated (and the temperature structure modified) by 
either (a) a reduction in the interstellar radiation field strength 
($G_0$), or (b) the inclusion of an external extinction ($A_{\rm v,ext}$) 
which would be appropriate if L1544 were surrounded by a 
low density background gas.
However, these are not degenerate; (a) would result in 
a reduction in the physical extent of the PDR, whilst (b) 
could result in the PDR being shifted so that it is completely outside 
the core. 

The initial conditions for the chemistry are taken to be atomic, with 
the H$:$H$_2$ ratio set to 10$^{-3}$ and the fractional abundance of 
CO, $X(CO)_0=10^{-6}$.
As explained above, a more realistic set of initial 
conditions could be obtained by evolving the chemistry to near-equilibrium.
Therefore, to test the sensitivity of the chemistry to the dynamical evolution 
prior to the intial dynamical configuration that we have adopted, we have 
included the possibility of a hydrodynamic `pause' 
($t_{\rm pause}$) at the start of the simulation. 
During this period the cloud is treated as being static and the physical 
conditions are held constant.

\section{Results}
\label{sec:results}

\subsection{The chemical structure}
\label{subsec:structure}

\begin{figure*}
\includegraphics[scale=0.8,trim=0 370 0 0,clip=true]{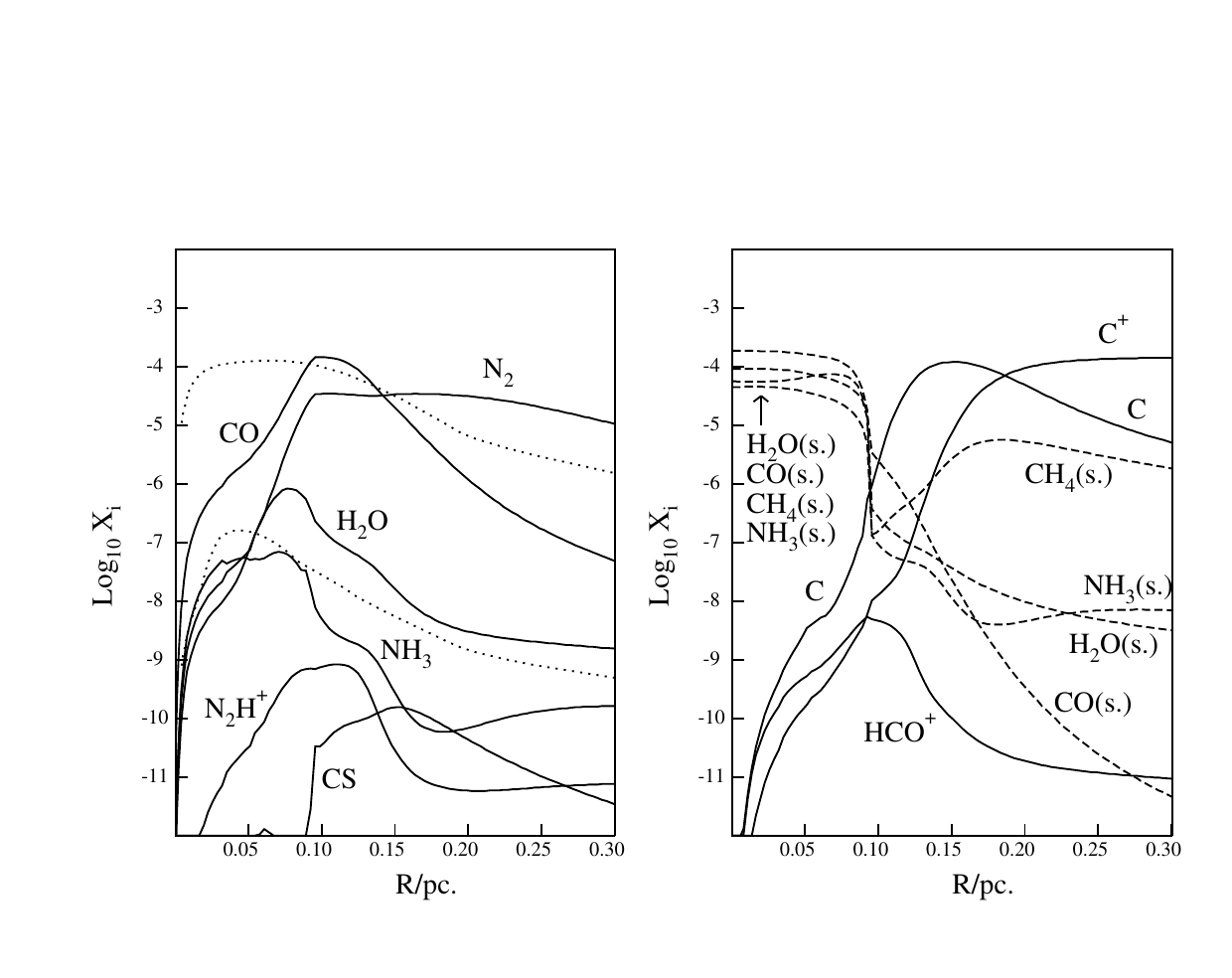}
\caption{Logarithmic fractional abundances (relative to H-nucleons) of 
several key species as a function of position at $t=t_{\rm max}=1.07$Myr.
Dashed lines are used for solid-state species, and the dotted lines 
show the abundance profiles for H$_2$O and CO, from KRC14, KC10 + KC08 
respectively.}
\label{fig:bestfit}
\end{figure*}

Results from the model are shown in Figure~\ref{fig:bestfit} for the values 
of the parameters given in Table~\ref{tab:param}, which are essentially
fixed for the specific chemical structure of L1544.
The figure shows the radial profiles of the fractional abundances of key 
species as functions of position at $t=t_{\rm max}$, i.e. corresponding to the 
current observational epoch. The fractional abundances 
are defined relative to the number density of hydrogen nucleons 
($n_{\rm nuc} = 2n_{\rm H_2} + n_{\rm H}$). This differs to the 
definition used in some other studies (e.g. KRC14), which is the relative 
to the H$_2$ number density. 

From various model calculations, and Figures~\ref{fig:physics} and 
\ref{fig:bestfit}, we find that it is possible to identify several different
chemical regimes.
From the outside, moving inwards:
\begin{enumerate}
\item 
For the gas densities applicable to the outer parts of the 
core, the transition between C$^+$/C/CO, characteristic of a photon 
dominated region (PDR) typically occurs at a mean A$_{\rm v}\sim 1-3$ 
magnitudes. In our model, this corresponds to a radial position of 
between $\sim 0.07-0.145$pc, where T$_{\rm dust}\sim 10.2-12.9$K.
Even at the outer boundary of 0.75pc ($A_{\rm v}\sim 0.07$), 
$T_{\rm dust}=17.8$K.
At these very low temperatures thermal desorption from dust grains is 
negligible for nearly all chemical species of interest, implying the 
existence of a `cold PDR' with a rather unusual gas-grain chemistry.
We identify two zones within the PDR boundary layer:
\begin{itemize}
\item the ice free zone; where $T_{\rm dust}\gtappeq 15$K 
($r\gtappeq 0.34$pc, $A_{\rm v}\ltappeq 0.28$). 
Here freeze-out and surface hydrogenation are inhibited by photodesorption, 
and the chemistry is dominated by gas-phase processes, and
\item the (marginal) ice formation zone; where $T_{\rm dust}\ltappeq 15$K,
freeze-out and surface hydrogenation occurs (yielding CH$_4$ 
and H$_2$O etc.), but is limited by photodesorption. 
Gas-phase chemistry still dominates in this region.
\end{itemize}
\item Interior to the PDR boundary layer there is the `dark core' with three
zones;
\begin{itemize}
\item an outer zone where the gas temperature falls rapidly, although the 
density does not rise as fast, so that freeze-out is once again inhibited,
\item an intermediate zone where the gas density rises significantly and 
freeze-out and photodesorption are both important; this is the location 
of the molecular (H$_2$O) abundance peaks, and  
\item an inner zone, where the density and extinction are high and 
freeze-out dominates \citep[in analogy with the static model of][]{Vas17}.
\end{itemize}
\end{enumerate}


The abundance profiles of the various species shown in Figure~\ref{fig:bestfit} 
are in broad agreement with those that are inferred from the L1544 observations.
For H$_2$O the principle empirical constraint on the gas-phase abundance 
profile comes from the radiative transfer analysis of the H$_2$O 567GHz 
($1_{10}-1_{11}$) line as observed by Herschel \citep{Cas12}.
The different chemical zones discussed above are clearly evident, with 
substantial ice mantle formation only occurring within the inner 
$\sim$0.08pc ($A_{\rm v}\gtappeq 2.6$).
Specifically, the H$_2$O abundance profile is consistent with what was 
presented in KRC14. 
The radial location of the abundance peak is shifted purely as a result 
of the use of an updated $A_{\rm v}$(r) profile.
The NH$_3$ abundance profile remains high until 
innermost depletion region ($r\ltappeq 0.02$pc, $A_{\rm v}\gtappeq 12$), 
consistent with observations \citep{Crap07,Cas22}.

\subsubsection{Ice composition}
\label{subsec:ices}

\begin{figure*}
\includegraphics[scale=0.8,trim=0 370 0 0,clip=true]{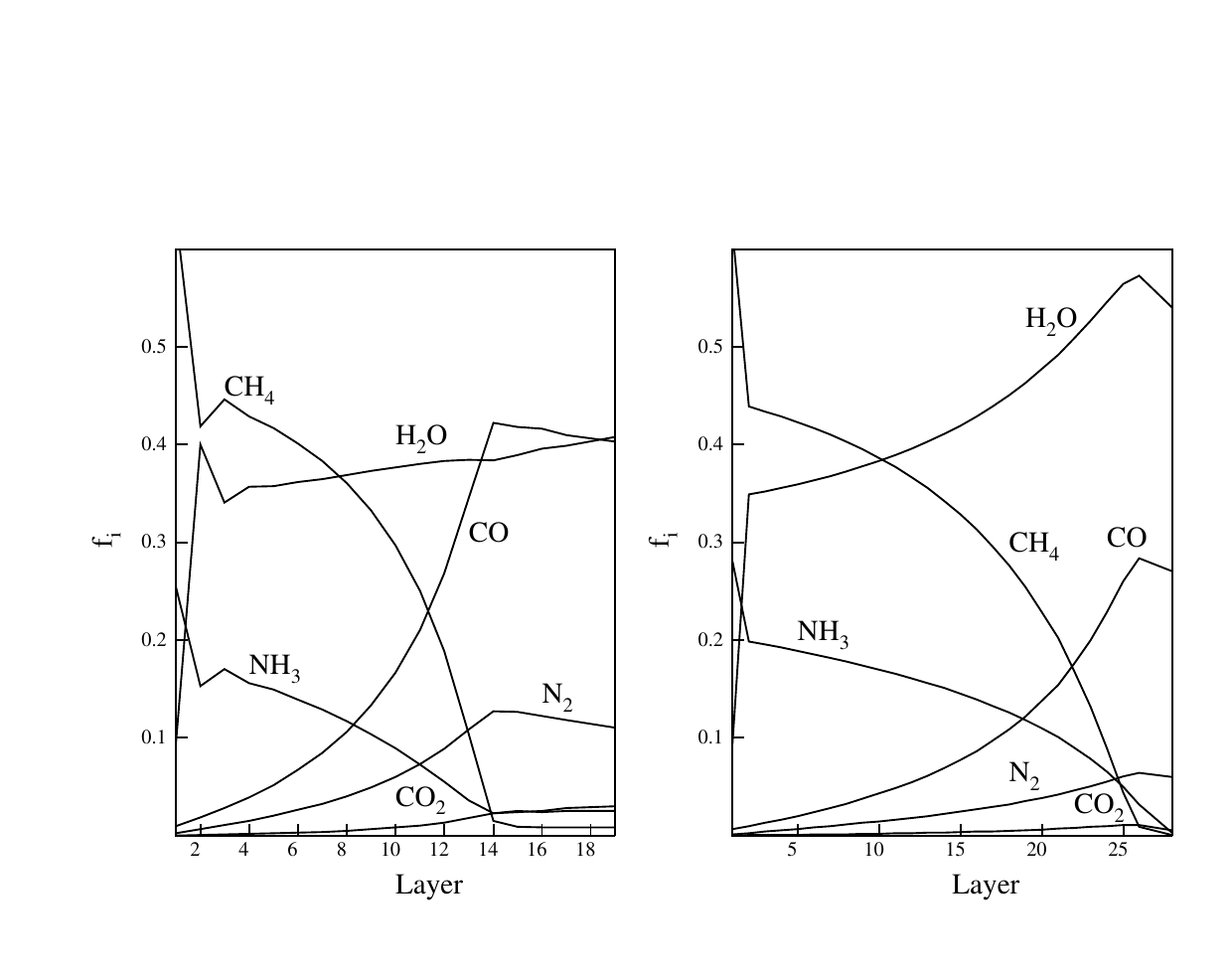}
\caption{Fractional composition of the ice mantle, as function of monolayer
number (with layer 1 on the lhs the lowest and the layer on the rhs corresponding 
to the surface of the ice mantle), 
for two positions at $t=t_{\rm max}$; r=0.08pc (lhs) and r=0.0005pc (rhs).}
\label{fig:ices}
\end{figure*}

In Figure~\ref{fig:ices} we show the compositional structure of the ices on the 
dust grains (in {\sc STARCHEM} represented by particles of a single size). 
The figure shows the fractional composition of
the ice for each monolayer, at two positions; near the outer edge of 
the ice formation zone (where photochemistry is important) and close to 
the (dark) centre of the core.
We can see that there are variations in the ice composition as a 
function of position in the core and that there are very significant 
stratifications within the ice mantles. 
In conditions where significant freeze-out occurs, the sub-surface layers of 
the ices - which are effectively chemically inert - are determined by the entire
evolutionary history of the cloud/core. 
Whilst, as we show in section~\ref{subsec:tdep}, the gas-phase abundances evolve 
quasi-statically, the ice has a long memory so that each monolayer retains the 
abundances that were in equilibrium with the gas phase abundances at their time 
of deposition. The compositional stratification of the ice mantles thus reflects 
the chemical evolution of the gas. The molecules H$_2$O and CO are seen 
to dominate the composition of the outer layers, as we would expect, because 
H$_2$O and CO are the dominant oxygen and carbon species inside the PDR. 
Moreover, the O and OH that freeze onto the grains are quickly hydrogenated to
H$_2$O. In contrast, in the inner layers of the ice mantles,
ammonia (NH$_3$) and, most especially, methane (CH$_4$) are major components.
The source of the CH$_4$ is incomplete conversion of C to CO in the earliest 
stages of the chemical evolution, resulting in the freeze-out and hydrogenation 
of C (see section~\ref{sec:free} below). 
A similar behaviour can explain the NH$_3$ and N$_2$ abundance 
stratification; the inner mantle samples the early time chemistry, when the gas 
has a large atomic component. Here, accreting N is converted to NH$_3$. 
The outer mantles sample the late time chemistry, when efficient conversion 
of N to N$_2$ has taken place, and there is little N in the gas-phase.

These results are quite different to those obtained by \citet{TCK12} with
the {\sc GRAINOBLE} model. This is due to a number of factors; \citet{TCK12} 
adopt a different grain size distribution, omit photodesorption and desorption
driven by the enthalpy of formation and - most significantly - they only included
the accretion of H, O and CO, so that the formation of CH$_4$ and NH$_3$ are
inhibited.

Unlike some studies \citep[e.g.][]{HC15}, we do not consider
the differences in binding energies for bare and icy grains, other than for the
desorption driven by the enthalpy of molecule formation. Whilst this has some
implications for the cosmic-ray heating-induced desorption of the most volatile 
species, this is generally not significant as the dust temperatures are so low
that thermal desorption is negligible for nearly all species of interest.

Our results can be used to compare with recent JWST observations of ice
mantle compositions toward dense dark clouds \citep[e.g.][]{McC23} and to guide 
future JWST observations of pre-stellar cores.

Small variations of the dust parameters (within the range of uncertainty)
can result in an
interesting phenomenon in the behaviour of the solid-state abundances of both 
CO and H$_2$O in the `cold core' region. These sometimes 
show strong enhancements in the range $0.18\ltappeq r\ltappeq 0.24$ pc
($0.44\ltappeq A_{\rm v}\ltappeq 0.7$).
Although the specific form of this feature is an artefact of the physical parameters
in our model (e.g. the simplifying assumption that dust grains can be represented by
a single size population), there is a genuine - if somewhat counter-intuitive - reason
for its existence: the sharp rise in the abundances at the outer positions correspond 
to the formation of ices of more than one monolayer thickness. However, whilst the 
desorption processes are only weakly dependent on local conditions, the freeze-out 
rates are proportional to the density ($n$) times the square root of the gas 
temperature ($T_{\rm gas}$). 
There is a region in the inflow where $T_{\rm gas}^{0.5}$ falls faster than $n$ rises 
and consequently the net freeze-out rate falls and the ices reduce to the 
sub-monolayer regime (within $r\ltappeq 0.18$pc, $A_{\rm v}\gtappeq 0.7$). 
In a real grain size population this enhancement would be smoothed out, but the 
overall effect is for dust grains to be relatively ice-free (i.e. in the 
sub-monolayer regime) right down to $r\sim$0.09pc ($A_{\rm v}\sim 2.2$).

\subsubsection{The chemistry of H$_2$O}
\label{subsec:H2Ochem}


To understand the behaviour of H$_2$O in more detail, we need to identify 
the dominant formation and destruction channels.
These have been discussed at some length 
elsewhere \citep[e.g.][]{DHN13} but to summarise; in the cold, dense environment 
of pre-stellar cores the predominant gas-phase formation route is driven by 
(cosmic-ray generated) H$_3^+$:
\[ {\rm O + H_3^+ \to OH^+ }\]
\[ {\rm OH^+ + H_2 \to H_2O^+ + H }\]
\[ {\rm H_2O^+ + H_2 \to H_3O^+ + H }\]
\[ {\rm H_3O^+ + e^- \to H_2O + H }\]
Neutral-neutral reactions, via the hydrogenation of the OH radical,
are only significant at higher temperatures, and
%
%
in cold environments the H$_2$O formation is 
dominated by surface reactions on interstellar ices. 
Several pathways have been identified \citep[e.g. see][]{DHN13,Iop08}, 
the most direct being the surface hydrogenation of accreted 
oxygen atoms and hydroxide (OH) radicals.
In addition, and particularly in relatively dense regions, significant 
contributions may be made via alternative 
(interconnected) channels involving O$_2$, either accreted from the gas-phase, 
where it is primarily formed
by the OH+O reaction, or else created via oxygen atoms reacting on the 
surface. The hydrogenation of surface O$_2$ leads to the formation of 
intermediate species en route to H$_2$O formation, such as 
HO$_2$ and H$_2$O$_2$. Alternatively, surface oxidation leads to the 
formation of O$_3$, which can be hydrogenated, again leading - ultimately 
- to H$_2$O formation \citep{CIRL10}.
Figure~\ref{fig:bestfit} shows different abundances in each
of the chemical zones identified in section~\ref{subsec:structure}; 
low molecular abundances are seen in the outer PDR region where 
photodissociation is active and also in the inner core where freeze-out
dominates. The molecular abundances peak at intermediate extinctions 
and densities.
In the outer regions ($r\gtappeq 0.2$pc, $A_{\rm v}\ltappeq 0.6$) most of the 
gas-phase H$_2$O derives from ice that has been desorbed by photodesorption 
and the enthalpy of surface formation of H$_2$O from O and OH. 
Indeed, if the latter process were suppressed, then the H$_2$O abundance would 
be lower by a factor of up to $\sim 10\times$.
The main H$_2$O loss mechanisms are photodissociation and reaction with C$^+$ 
(to form HCO$^+$, ultimately leading to CO formation). 
This effectively inhibits ice formation until $r\ltappeq 0.1$pc 
($A_{\rm v}\gtappeq 1.9$).
%
%
At the location of the H$_2$O abundance peak ($\sim 0.08$pc, 
$A_{\rm v}\sim 2.6$) photodesorption and formation enthalpy-driven desorption
still dominates the production of gas-phase H$_2$O, although gas-phase 
reactions (via H$_3$O$^+$) also contribute significantly. 
Photodissociation is the dominant destruction mechanism.
In the inner core cosmic-ray induced photodesorption dominates the production 
of gas-phase H$_2$O. As the rates for this are rather 
ill-defined there is a considerable margin of error in the model predictions.
In the innermost regions ($r\ltappeq 0.02$pc, $A_{\rm v}\gtappeq 12$) the 
abundance declines due to efficient freeze-out at high densities - but gas-phase 
destruction by H$_3^+$ accounts for $\sim 30$\% of the loss rate. 
The surface hydrogenation of accreted OH is the main formation channel.
%


\subsubsection{The chemistry of CO}
\label{subsec:COchem}

Unlike H$_2$O, which is mainly formed on the surface of grains, CO
is formed in the gas-phase.
In dense cloud conditions this chemistry is relatively slow, 
compared to other gas-phase processes and freeze-out, but may result in 
the efficient conversion of C to CO.
Several channels are possible, including:
\begin{equation}
{\rm C^+ + H_2 \to CH_2^+ + h\nu }
\label{eqn:CPH2}
\end{equation}
\begin{equation}
{\rm CH_2^+ + O \to HCO^+ + H }
\label{eqn:CH2PO}
\end{equation}
\begin{equation}
  {\rm HCO^+ + e^- \to CO + H }
\label{eqn:HCOPrec}
\end{equation}
or
\[ {\rm C + H_2 \to CH_2 + h\nu }\]
\[ {\rm CH_2 + O \to CO + 2H }\]

However, the initiating radiative association reactions for both of these channels
are very slow;
the reactions of C and C$^+$ with H$_2$ to form CH and CH$^+$ have large
activation barriers, whereas the corresponding radiative association reactions
(to form CH$_2$ and CH$_2^+$) do not possess inhibitive barriers, but are inefficient.
In normal dark cloud environments, reactions involving OH may therefore be
more important:
\begin{equation}
{\rm C + OH \to CO + H }
\label{eqn:COH}
\end{equation}
or
\begin{equation}
{\rm C^+ + OH \to CO^+ + H }
\label{eqn:CPOH}
\end{equation}
\begin{equation}
  {\rm CO^+ + H_2 \to HCO^+ + H }
\label{eqn:HCOP}
\end{equation}
followed by reaction (\ref{eqn:HCOPrec}),
although the efficacy of reactions~(\ref{eqn:CPOH}) and (\ref{eqn:HCOP}) have 
recently been challenged by the non-detection of CO$^+$ in diffuse 
interstellar clouds \citep{GL21}, which suggests that non-thermal reactions involving
C$^+$, CH$^+$, CH$_2^+$ etc. and oxygen atoms are likely to be a more 
important source of the observed HCO$^+$:
\[ {\rm C^+ + H_2 \to CH^+ + H }\]
\[ {\rm CH^+ + H_2 ...\to CH_2^+,CH_3^+... + O \to HCO^+ + ... }\]
But, as noted above, the first of these reactions is very slow at low 
temperatures. 

Interior to the PDR, at the densities applicable to pre-stellar cores, the 
conversion of C to CO goes to near completion (i.e. $X(CO)\sim X(C_{\rm total})$)
within a few$\times 10^5$ years
and the ratio of the gas-phase to solid-state CO abundances is set by 
the balance of freeze-out and desorption.
CO has a much lower surface binding energy than H$_2$O so that the 
gas-phase abundance is sensitive to the assumptions concerning the
desorption efficiencies. 
Thus, although the dust temperature is low enough that thermal desorption
is inhibited throughout this region, desorption by cosmic-ray heating
is important (and dominates over photodesorption). The 
effect of chemical desorption (driven by H$_2$ formation) may also be 
significant \citep[e.g.][]{Pant21}.

Further out, within the PDR zone, CO formation is incomplete and the 
abundances of C and C$^+$ are relatively high. Indeed, even deep inside 
a PDR, the abundances of C and C$^+$, do not vanish and are typically 
of the order of $10^{-4}-10^{-3}$ times that of CO 
\citep[e.g. see Fig 7 of][]{Retal07}.
The situation in a `cold PDR' is more complex; in (the oxygen rich) 
gas-phase, carbon chemistry tends to lead to the formation of the stable CO 
molecule as the dominant product (leaving free volatile oxygen to form 
H$_2$O, O$_2$ and other species), with a much smaller proportion of
the available carbon present in other organic molecules.
However, on the surface of grains, it is more likely that any accreting
carbon atoms or ions are hydrogenated to form methane (CH$_4$).

If the conditions are propitious for freeze-out (as is 
the case in the cold PDR in the outer regions of the core), then there will 
be a competition between 
\[ {\rm C ~or~ C^+ + H_2 ~or~ OH ...\to CO  }\]
\[ {\rm CO + grain \to CO(s) }\]
also leading to methanol (CH$_3$OH) formation (which we do not discuss
in this study; see Appendix~\ref{sec:starchem}), and
\[ {\rm C ~or~ C^+ + grain \to CH_4(s) }\]
The very low temperatures and the chemical structure of pre-stellar cores 
imply that the gas-phase and gas-grain chemistries pull in different 
directions (leading to CO and CH$_4$ respectively).
This implies that the efficiency of CO formation in the pre-contraction phase 
is important in determining the balance between CO/CH$_3$OH and CH$_4$.
This balance will be strongly dependent
on the assumptions concerning the grain size distribution. Thus, if the 
grain surface area is predominantly carried by negatively charged very small 
grains (as is the case in the MRN dust size distribution), then there could be 
a strong electrostatic enhancement of the freeze-out of C$^+$, favouring the 
methane formation channel. 
We defer a full discussion of the importance of the grain size distribution 
to a future study.

Referring to Figure~\ref{fig:bestfit} we can see that 
the qualitative behaviour of the CO abundance profile is similar to that for
H$_2$O. That is to say there are essentially three chemical regimes;
(i) an outer PDR zone, where $X(CO)$ tails off at low A$_{\rm v}$,
(ii) a middle region where $X(CO)$ reaches near saturation, and
(iii) an inner depletion zone where the depletion factor depends on the ratio
of the freeze-out to the cosmic-ray induced desorption rates.

We confirm the findings of KC10; that an efficient desorption mechanism is
{\em required} to maintain sufficiently high CO abundances at high 
extinctions, to explain the observed line strength. 
The most efficient mechanism for CO desorption is cosmic-ray grain heating - 
but, as deduced by KC10 from a comparison with the observed brightness of CO
in L1544, the rate would need to be $\sim 10-30\times$ larger than given 
in \citet{HH93} \footnote{The uncertainty of a factor of three in this estimate
comes from the non-linear scaling of the CO brightness with the cosmic-ray 
desorption, the accuracy of the radiative transfer modelling and the 
observational uncertainty}.
Current studies \citep{Raw22} indicate that cosmic-ray heating induced 
desorption rates may have been grossly underestimated in previous studies
\citep[see also][]{Sil21}.
There are, however, very significant uncertainties in the quantification of
these rates and their $A_{\rm v}$-dependence but these, empirically deduced
values, are within this range of uncertainty.

In our model we find that the formation of CO is a little more complex than 
H$_2$O; in the outer regions several 
formation channels are effective, including dissociative recombination of
HCO$^+$, but also via the photodissociation/hydrogen abstraction of 
surface-formed CH$_4$, yielding CH, which reacts with oxygen atoms. 
A corollary of this pathway multiplicity is that the CO abundance profile is 
fairly robust to uncertainties/variations in the rate coefficients for specific
reaction pathways.
The destruction of CO is completely dominated by photodissociation (in
the PDR) and freeze-out (in the inner core). At intermediate radii
(0.08pc$\ltappeq r \ltappeq 0.15$pc, $0.96\ltappeq A_{\rm v}\ltappeq 2.6$) 
reaction with H$_3^+$ to form HCO$^+$ is 
significant, although most of the HCO$^+$ is re-cycled to CO.
In the innermost regions, there is little gas-phase chemistry and the abundance
is essentially established by the balance between (cosmic-ray heating) desorption
and freeze-out.

\subsubsection{CO and the nitrogen chemistry}
\label{subsec:nitro}

Some previous studies \citep[e.g.][]{Holl09} have predicted gas-phase CO to be 
concentrated in a `ring' at A$_{\rm v}\sim1$.
Part of the reason for this is due to the coupling of the chemistries for 
CO and H$_2$O (e.g. as is evident from reactions \ref{eqn:COH} and 
\ref{eqn:CPOH} above). But, most significantly, it was postulated that in 
the regions where H$_2$O freezes out but (the more volatile) CO does not, CO
is destroyed (by He$^+$, N$_2$H$^+$ and photodissociation at low $A_{\rm v}$) 
yielding O.
This can then freeze out and be converted to H$_2$O on grains,
so that the net result is a gradual shift of the oxygen from CO 
to H$_2$O, and an anticorrelation between CO and N$_2$H$^+$.
Our model results do not show this anticorrelation. Observationally,
there is some evidence of a rise in the abundance of 
N$_2$H$^+$ in high density regions where CO is depleted 
\citep[e.g.][]{Cas99,Aik01}, although this observation could be an artefact of
radiative transfer effects \citep{Cas22}. 
We defer a full description of the chemistry of N$_2$H$^+$ to a future
study, but our results indicate that the abundance profiles of CO and N$_2$H$^+$ 
are only weakly connected.
We also find that destruction by He$^+$ is insignificant at all positions. 

The nitrogen chemistry is fairly simple:
throughout much of the core, the chemistry of N$_2$ is driven by CH:
\[ {\rm  CH + N \to CN} \]
\[ {\rm CN + N \to N_2} \]
and, here, reactions with He$^+$ are the dominant destruction mechanisms 
(ultimately providing formation channels for NH$_3$ and N$_2$H$^+$), although 
reaction of N$_2$ with H$_3^+$ to form N$_2$H$^+$ is significant at the 
innermost positions.
In these regions of high depletion, the hydrogenation of N$^+$ to NH$_4^+$, 
followed by dissociative recombination, is also a major formation 
channel for NH$_3$.
As with CO the dominant desorption mechanism for other volatile species, 
such as N$_2$, is found to be direct cosmic ray heating of grains.
%

\subsection{Time-dependence}
\label{subsec:tdep}

\begin{figure}
\includegraphics[width=0.93\columnwidth,trim=0.7in 5.7in 4.1in 1in]{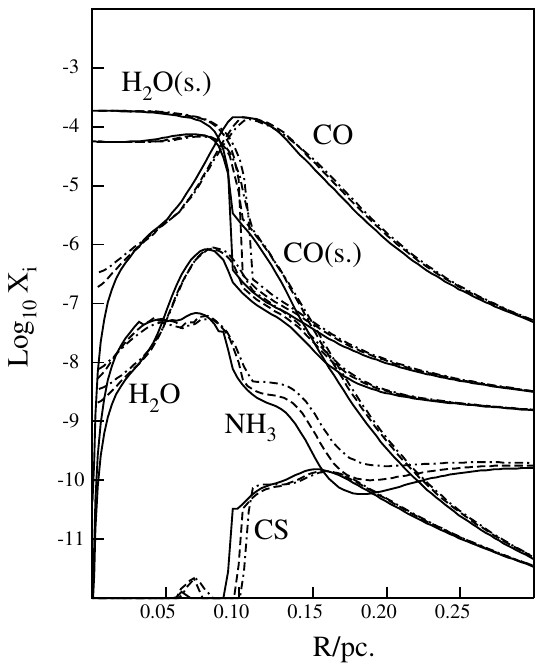}
\caption{Logarithmic fractional abundances (relative to H-nucleons) of 
selected molecular species as a function of position at three times:
t$_{\rm min}$=0.30Myr (dot-dashed lines), 0.45Myr (dashed lines), 
and t$_{\rm max}$=0.61Myr (solid lines).}
\label{fig:timedep}
\end{figure}

In Figure~\ref{fig:timedep} we show the radial abundance profiles for selected
species (H$_2$O, H$_2$O(s), CO, CO(s), NH$_3$ and CS) obtained at three 
specific times in the dynamical/chemical evolution: 
t$_{\rm min}$=0.30Myr, 0.45Myr and t$_{\rm max}$=0.61 Myr. 

We can immediately see from this figure that there is hardly any `intrinsic' 
time-dependence of the chemical abundances. The abundances do change with 
position and time, particularly in the inner regions, but this is primarily a 
result of the contraction of the core and the adjustment of the chemical 
equilibrium to the changing physical conditions. 
To test this result we have re-calculated these profiles, but 
setting $t_{\rm pause}=0.5$Myr - i.e. inserting a period of chemical 
evolution, during which the physical parameters are held constant, at the
beginning of the simulation. This yields almost identical results to those
shown in Figure~\ref{fig:timedep}, other than a small reduction (by a factor of
$\ltappeq 2$ in the abundance of NH$_3$ in the range 0.15pc$<r<$0.25pc,
$0.42\ltappeq A_{\rm v}\ltappeq 0.96$).
This result indicates that it would not be possible to determine a 
`chemical age' for the cloud.
Although the chemical evolution 
is in quasi-equilibrium, in the parts of the core where thick ice mantles are 
present it is possible that some abundances will be affected by the prior
evolution of the core. This is because the equilibrium is determined by the balance
between gas-phase and gas-surface interactions. In our model the previously 
accreted sub-surface ice layers are assumed to be inert and effectively removed 
from the active chemistry.

With the assumption of quasi-equilibrium, we can estimate the gas-phase 
abundance of H$_2$O in the inner parts of the core as balance between freeze-out 
and desorption. We can also calculate the relative contribution from each of
the applicable desorption mechanisms; cosmic ray induced photodesorption, cosmic ray
heating induced desorption and H$_2$-formation induced desorption.

Comparing the desorption efficiency for cosmic-ray induced photodesorption
with that for H$_2$ formation-induced desorption, we find that the two are 
equal when
\[ {\rm\left(\frac{Y_{H_2}}{Y_{pd}}\right) = \left(\frac{0.85}{n_H}\right)
\left(\frac{\zeta}{1.3\times 10^{-17}s^{-1}}\right). }\]
Thus, in normal circumstances, this implies that values of 
$Y_{\rm H_2}\gtappeq 10^{-3}$ would be needed for H$_2$ formation-induced 
desorption to be important.

Then, comparing the desorption efficiencies for cosmic ray heating-induced desorption 
with cosmic ray induced photodesorption we find that the former dominates when 
\[ {\rm  k_{cr} > \left( \frac{Y_{pd}F_{cr}}{N_s}\right) 
\gtappeq 5\times10^{-15}s^{-1} }\]
where we have used parameter values as given in Table~\ref{tab:param}.
For species of interest, most estimates of $k_{\rm cr}$ are less than this value, 
but we cannot exclude the possibility that recent revisions could imply that 
cosmic ray heating may the dominant desorption mechanism for most species in dark
environments.

Hence, balancing freeze-out with cosmic-ray induced photodesorption, and with the 
additional assumptions that $T_{\rm gas}=10$K and that the proportion of surface
coverage of grains by water molecules is of order unity (as indicated in 
Figure~\ref{fig:ices}), then - again using parameter values from 
Table~\ref{tab:param} - the equilibrium gas-phase density of H$_2$O is
\[ {\rm n_{H_2O} = 1.8\times 10^{-3}\left( \frac{Y_{pd}}{10^{-3}}\right)
\left( \frac{\zeta}{1.3\times 10^{-17}s^{-1}}\right) cm^{-3}}.\]
Assuming that $Y_{\rm pd}$ and $\zeta$ do not change as functions of 
position, this is independent of the local conditions, so that the fractional
abundance of water, $X(H_2O)$, simply scales as the inverse of the density.
This behaviour can be seen in the inner regions ($r\ltappeq 0.05$pc, 
$A_{\rm v}\gtappeq 4.8$) in Fig~\ref{fig:timedep}. 

\subsubsection{Parameter variations}
\label{sec:free}

We have also investigated the sensitivity of these results to the values 
of some of the parameters (identified in Table~\ref{tab:param}), the chemical 
initial conditions and the choice of reaction database, within the 
ranges that are appropriate for pre-stellar cores. 
We defer a full discussion to a future paper but highlight some key
results here.

Apart from some obvious dependencies on the cosmic ray induced desorption
rates and the various desorption efficiencies (particularly in the 
inner core) the H$_2$O and CO abundance profiles are, in general, found 
to be fairly insensitive to most of these parameter variations. For most
species, a range of gas-phase, gas-grain and surface reactions operate at 
all positions within the core.
Changing the initial conditions (specifically, the initial fractional 
abundance of CO) has little effect on the gas-phase abundances - 
consistent with the condition of quasi-static equilibrium - but 
results in significant variations in the CH$_4$ to CO ratio in the ice 
mantles. This is an interesting result that will be investigated in more 
detail in future studies. 

The chemistry is, however, very sensitive to the value of the external 
extinction, as this affects both the radiation field strength and
the location and extent of the PDR. We find that the inclusion of even 
modest values of $A_{\rm v,ext.}\sim 1-2$ results in a shift in 
the location of the H$_2$O abundance peak and large abundances in the 
outer envelope that are quite incompatible with the empirically
constrained abundance profile. 
There is filamentary structure in the vicinity of L1544, and the presence 
of a parsec-scale low density ($\sim 30$cm$^{-3}$) medium is required to 
reproduce the HCO$^+$(1-0) observations \citep[e.g.][]{Red22, Giers23}.
However, the regions that are responsible for the observed CO and H$_2$O 
line profiles cannot be obscured from the interstellar radiation field 
by material with significant extinction in the near-IR and UV.

\subsection{The applicability of reduced chemical networks}
\label{sec:lite}

In this section we compare the accuracy of using a highly reduced chemical 
reaction network relative to a full description of the gas-grain chemistry.

In a cold, dark cloud most of the oxygen budget is taken up by O, O$_2$, 
CO, OH and H$_2$O. As CO is relatively chemically inert and O$_2$ is not a 
dominant component, KRC14 proposed that the H$_2$O 
chemistry could be described adequately by a highly reduced network of just
a few reactions. In this scheme the gas-phase chemistry is limited to 
the hydrogenation and photodissociation reactions:
\[ {\rm O + H_2 \to OH + H} \]
\[ {\rm OH + H_2 \to H_2O + H} \]
\[ {\rm H_2O + h\nu \to OH + H} \]
\[ {\rm OH + h\nu \to O + H} \]
Although the hydrogenation reactions are vanishingly slow at temperatures 
below a few hundred degrees, they are included to provide reverse reactions 
for detailed balance with the photodissociation reactions. With these reactions, 
the  reduced chemical network is also applicable in warmer gas.

The gas-grain interactions simply include the freeze-out
and (assumed complete) surface hydrogenation of O,OH (and H$_2$O):
\[ {\rm O, OH, H_2O \to H_2O_{(ice)}} \]
balanced by desorption:
\[ {\rm H_2O_{(ice)} \to OH~and~H_2O} \]
The processes include direct 
cosmic ray heating, photodesorption and (in KRC14) cosmic-ray 
induced photodesorption. Thermal desorption, though included, is 
insignificant in starless cores because of the low dust temperatures.

KC08 and KC10 assume that the carbon chemistry could be decoupled 
from the oxygen/H$_2$O chemistry and be reduced to just three reactions:
\[ {\rm C + h\nu \to C^+ + e^-} \]
\[ {\rm CO + h\nu \to C + O} \]
and the (slow) radiative association of C$^+$ with H$_2$ 
(reaction~\ref{eqn:CPH2}), which is the rate-limiting step for CO formation 
(followed by reactions \ref{eqn:CH2PO} and \ref{eqn:HCOPrec}). 
The only gas-grain processes were taken to be the freeze-out and (cosmic 
ray-induced) desorption of CO.

%

\subsubsection{Comparison of reduced with full chemical networks} 
\label{subsec:compare}

As mentioned in the introduction, one advantage of the reduced chemical networks 
is the simple and rapid calculation of the abundances of the dominant coolants 
using a few algebraic equations. These apply to each of the H$_2$O and CO 
chemistries and relate the abundances to the values of the model-dependent 
reaction rates. This allows for a simple analytical approximation to establish 
the temperature of the gas \citep{Gold01,KF05} which is valid even across PDR boundaries.

Previously we compared our dynamical and chemical model of L1544 to observations 
of L1544 (KC10, KRC14). The predicted C$^{18}$O and H$_2$O abundance profiles for 
L1544 along with radiative transfer modelling produced C$^{16}$O, C$^{17}$O and 
C$^{18}$O (1-0) and H$_2$O 567GHz (1$_{10}-1_{11}$) line profile shapes and 
strengths consistent with those observed. The reduced chemical 
network is also responsible for the gas temperature which factors 
into the line strengths. In this sense the method is observationally verified. 

A direct comparison of the abundances produced by the reduced chemical network 
with those from a full network has the advantage that the accuracy of the 
approximation can be determined at least for the examples compared. 
To assess the accuracy of the reduced network chemistry, we have compared the 
abundances at $t_{\rm max}$ from our time-dependent full network chemical model 
with the equilibrium abundances at $t_{\rm max}$ also from STARCHEM but including 
only the reactions from the reduced network as defined above. 
This effectively mimics the analytical approach used in our earlier papers, 
KC08, KC10, and KRC14.

The results from this comparison show that whilst there is a good fit to the 
H$_2$O abundance profile, the fit for the CO profile is not as good; the 
reduced network results in too much CO in the outer parts (due 
to the omission of various gas-phase destruction channels) and too little CO
in the inner regions (due to the slowness of the CO production reactions).
The C and C$^+$ profiles are also problematic, with high abundances persisting
down to the centre of the core. This is again partly due to the slowness 
of the C to CO conversion reactions in the reduced network.
 
An obvious problem with the carbon chemistry in the reduced network as proposed by 
KC08 
is that C$^+$ recombination was not included:
\[ {\rm C^+ + e^- \to C + h\nu} \] 
and, other than photoionization, there is no loss channel for carbon atoms.
We have included additional channels in the network to account for the conversion 
of C and C$^+$ to CO (as suggested in section~\ref{subsec:COchem}):
\[ {\rm C + OH \to CO + H} \]
\[ {\rm C^+ + OH \to CO^+ \ldots\to CO} \]
(assuming that every CO$^+$ molecule so-formed leads to CO formation, via 
reactions \ref{eqn:HCOP} and \ref{eqn:HCOPrec}).
Here, the abundance for OH is found from the reduced network for oxygen. 
The loss of OH in these reactions should not much affect the abundances in the 
oxygen network since the total abundance of oxygen is three times that of carbon, 
and we ignore this effect.

\begin{figure*}
\includegraphics[scale=0.8,trim=0 370 0 0,clip=true]{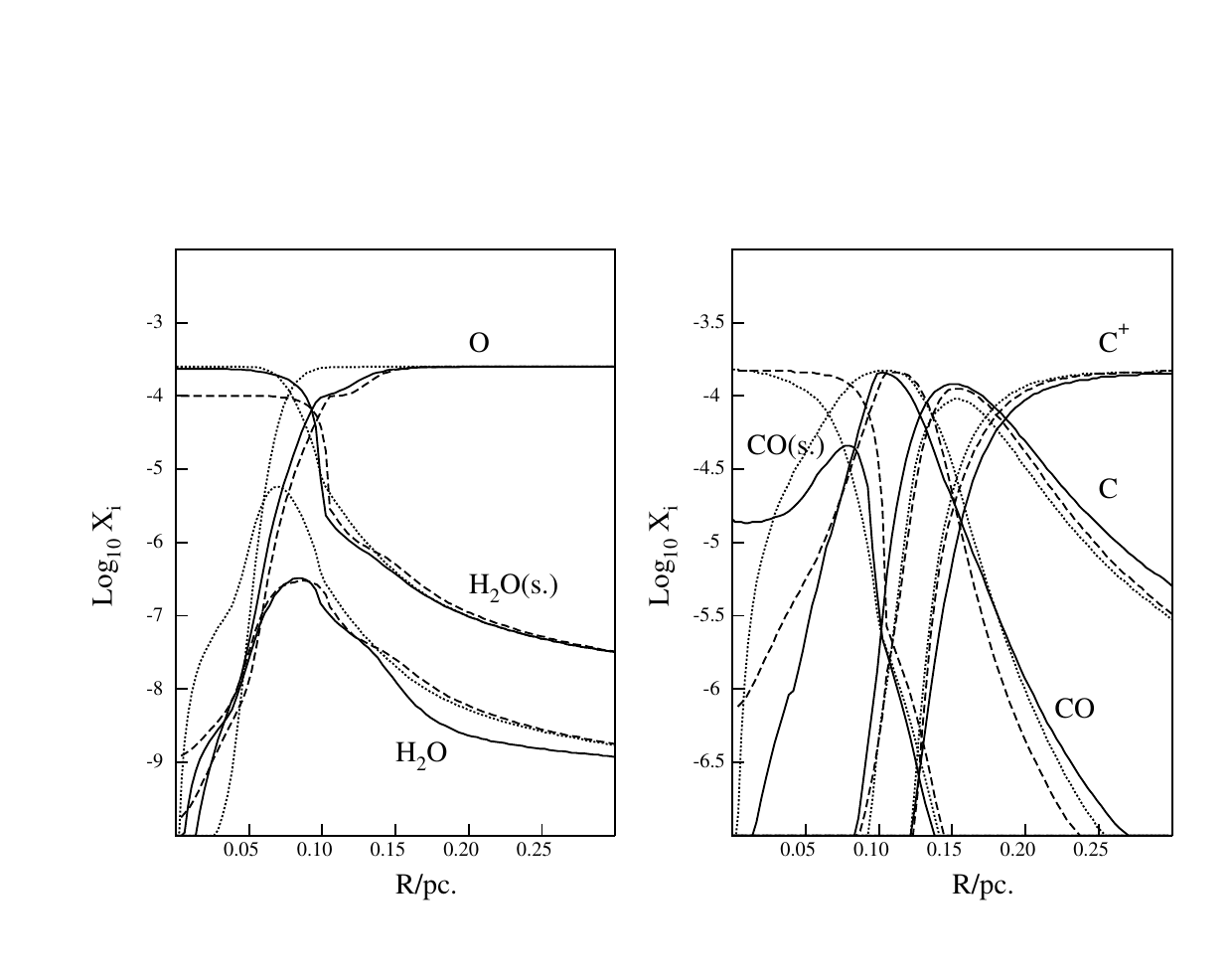}
\caption{Radial abundance profiles of H$_2$O, H$_2$O(s), O, C, C$^+$, CO and CO(s.)
(at $t=t_{\rm max}$). Results are shown for the full chemical/dynamical
model (solid lines), the revised reduced network chemistry evolved to equilibrium 
(dashed lines), and the solutions of the algebraic equations (dotted lines).}
\label{fig:lite}
\end{figure*}

In Figure~\ref{fig:lite} we show the results of a comparison between 
the full network and our reduced networks for the H$_2$O chemistry
(as in KRC14) and the CO chemistry (revised, as described in 
Appendix~\ref{sec:CO_lite}).
The figure shows the abundance profiles (at $t=t_{\rm max}$)
obtained for H$_2$O, H$_2$O(s), O, C, C$^+$, CO and CO(s.).
To compare `like with like' we suppress enthalpy-driven desorption and the 
surface chemistry of CO (leading to CO$_2$ formation etc.) in these 
simulations.
In this figure, we also show the abundances determined directly from the 
algebraic equations for chemical equilibrium with the reduced networks.

Using the revised reduced networks in the full model yields remarkably 
accurate results (typically within a factor of $\sim 2\times$, at all radii) 
for all species, other than CO and CO(s.) in the innermost region where 
there is some discrepancy. This is due to the large methane abundance 
obtained in this partcular model, which is not accounted for in the reduced
network.

The algebraic solution also gives excellent fits, except for CO and 
H$_2$O which are over-estimated in the inner regions, where the dust grains
have thick ice mantles. The reason for this is straightforward; the desorption 
rate of a species depends on the fraction of the grain surface that it covers.
The algebraic formulae take the desorption rate to be proportional to the total
ice mantle abundance of that species. In the sub-monolayer regime this is a
satisfactory approximation, but for thicker ices this over-estimates the 
desorption rate by a factor of order of the number of monolayers. 
This discrepancy has only a marginal effect on the ice abundances as they
are much greater than the gas-phase abundances in these conditions.

We conclude that the reduced networks (including the modifications for the CO
chemistry as described in Appendix~\ref{sec:CO_lite}) work extremely well, 
although their accuracy in regions of high depletion are contigent on 
H$_2$O and CO being the dominant ice mantle components.
We recommend that the revised CO chemistry network be used in future 
applications.
The algebraic solutions to the networks also work well, except for CO and 
H$_2$O in regions of high depletion.

\subsubsection{Utility of reduced network chemistries} 
\label{subsec:utility}

In pre-stellar cores the main heating mechanism for the gas throughout most of 
the core is cosmic ray heating.
In the outer region, the UV flux on dust also heats the gas through the 
photoelectric effect. In the central region which is completely shielded from 
starlight, only cosmic ray heating is effective, and the gas becomes very cold, 
approaching 5K. 

The main coolants for the atomic gas are [CII] and [OI] fine structure lines, 
whilst the molecular gas cools primarily through the CO rotational lines and 
dust thermal emission. The 
most effective optical depth for cooling is about unity. Where the CO(1-0) is 
optically thick, the higher excitation lines become effective coolants as well 
as a few less abundant simple carbon species \citep{Gold01}. 
Figure~\ref{fig:physics} shows a local maximum in the gas temperature at a 
radius of 0.03 pc. This is due to the variation of the optical depths of the 
molecular line coolants, primarily CO(1-0). As the dust temperature increases 
with radius due to the increasing UV flux, so does the gas temperature by 
collisional coupling with the dust. However, at radii $\sim$0.03 pc, the 
optical depth to the boundary of the cloud decreases enough that the cooling 
through the molecular lines becomes more effective. At radii $>$ 0.3 pc, the 
gas temperature decreases despite the steady increase in dust temperature. 
The rapid rise in the gas temperature at the boundary is caused by hot 
electrons photoelectrically released from dust grains by FUV radiation that 
penetrates only a short distance into the cloud.

\section{Summary and conclusions}
\label{sec:conc}

We have applied an empirically constrained multi-point dynamical model, coupled 
to a comprehensive description of the chemistry, to a study of the chemical 
processes and timescales in pre-stellar cores. We have applied this model to
L1544 to study the temporal and spatial evolution of the abundances of H$_2$O, 
CO and other key small molecular species.

The key results from our studies are as follows:
\begin{enumerate}

\item
At the time corresponding to the epoch of observation of L1544, the chemical 
timescales are less than the dynamical timescale throughout the core. Hence, the
chemistry is in near equilibrium; that is to say, the chemical abundances at 
any position simply respond to the instantaneous local physical conditions 
($n,T,A_{\rm v}$) and are, to a large extent, independent of the prior chemical 
and dynamical evolution. 
The H$_2$O abundance profile in this type of source only depends on the 
physical structure, and the values of certain chemical parameters.
This is a significantly different result to what is usually assumed to be the
case for young star-forming regions and implies that, for isolated pre-stellar 
cores at least, the chemical and dynamical 
evolution are not separable. {\bf Hence, the concept of a 
chemical (or depletion) `age', and the applicability of `chemical clocks',
are meaningless}.
The only situations in which chemical clocks may be relevant are the 
highly unusual conditions where an observed system has assembled on a 
timescale that is short compared to the chemical timescale, but then evolves 
on a very much longer timescale.

\item 
The ice mantles on dust grains are highly structured, so that whilst 
CO and H$_2$O dominate the outer layers other species, such as CH$_4$ and 
NH$_3$, are important in the inner layers. 
As has been previously noted \citep[e.g.][]{TCK12,RWW19}, 
if it is assumed that the sub-surface ice layers are chemically inactive 
they act as a fossil record/reservoir of the prior dynamical/chemical evolution.
This implies that the quasi-equilibrium gas-phase abundances are not 
completely decoupled from the evolutionary history of the core.
For example, the degree of conversion of C to CO in the earliest 
evolutionary stages determines the overall abundance of CH$_4$ ice (due to
the competition between gas-phase chemistry; yielding CO, and surface chemistry;
yielding CH$_4$). 
A similar result applies to NH$_3$ ice, whose abundance is determined by
the degree of conversion of N to N$_2$ in the gas-phase.

\item The model matches the empirically determined abundance profiles for 
H$_2$O and CO in L1544. 
The chemistry of hydrogenated species, such as H$_2$O, NH$_3$ 
and CH$_4$, is dominated by surface reactions. For many other species there
are gas-phase and surface chemistry formation channels that operate
throughout the range of physical conditions that exist within the core, so 
that their abundance profiles are remarkably robust to parameter variations.

\item The highly simplified reduced network chemistries for oxygen species 
described in KRC14, and for C/C$^+$/CO, described in KC08 \& KC10 and revised 
here (Appendix~\ref{sec:CO_lite}), give a very accurate description of the 
chemistry and abundance profiles for gas-phase and solid-state H$_2$O and CO, 
OH, O, C and C$^+$. This is valid over the whole range of extinction 
(1$<$A$_{\rm v}<$1000) applicable to starless cores and quiescent molecular gas. 
The accuracy of the CO and H$_2$O profiles in regions of high depletion depends 
on these species being the dominant components of the ice mantles.
Algebraic solutions to these networks also give good descriptions of the abundance
profiles, but over-estimate the gas-phase abundances of CO and H$_2$O when 
significant freeze-out has occurred.
%
%
%

\item The abundance profiles for both H$_2$O and CO in L1544 necessitate the 
existence of a PDR in the outer regions of the core and the interstellar
radiation field must reach the outer layers without attenuation 
(i.e. $A_{\rm v,ext}\sim 0$). This is indicative of the presence of highly 
non-uniform density and extinction structures in this source.

\item The chemical structure is quite complex, with both gas-phase and
surface chemistry important throughout the core. In the outer regions `cold PDR'
conditions prevail, with photodesorption and enthalpy-of-formation driven 
desorption dominating the gas/ice balance.
Thermal desorption is only significant in the outermost parts of the core,
and even there is only limited to highly volatile species. 

\item The chemistry in the inner (dark) regions of the core 
($r\ltappeq 0.08$pc, $A_{\rm v}\gtappeq 2.6$) is dominated by 
cosmic-ray induced processes. In particular, 
the desorption of ices by cosmic ray induced photodesorption (for H$_2$O) and 
cosmic ray grain heating (for CO etc.) are the main factors that determine the 
gas-phase abundances in the highly depleted regions.
However, in the light of recent re-calculations \citep[e.g.][]{Raw22}, we find 
that in these conditions, desorption driven by the cosmic ray heating of grains 
is probably more significant than cosmic-ray induced photodesorption
for a number of volatile species, including CH$_4$ and N$_2$.

\end{enumerate}

Future modelling efforts will include updates to include a more 
comprehensive description of the (position-dependent) grain size 
distribution and its several effects on the chemistry. 
This update would also include allowances for the ice 
composition-dependences of the binding energies and desorption 
efficiencies.

Our modelling of the H$_2$O and CO abundance profiles in 
a specific pre-stellar core (L1544) can be extended to study the 
abundance profiles of, and correlations between, other key molecular 
tracers, such as N$_2$H$^+$, N$_2$, HCO$^+$, H$_2$CO and CH$_3$OH.
The flexibility of the model is such that this could easily be adapted 
and applied to studies of other pre-stellar cores.

\section*{Data availability}

The data underlying this study are openly available from the published
papers that are cited in the article.
The data generated in support of this research are partly available
in the article, and will be shared on reasonable request to the corresponding
author.



\bsp


\appendix
\section{Simplified chemical networks}
\label{sec:CO_lite}

\begin{figure}
\includegraphics[scale=0.3]{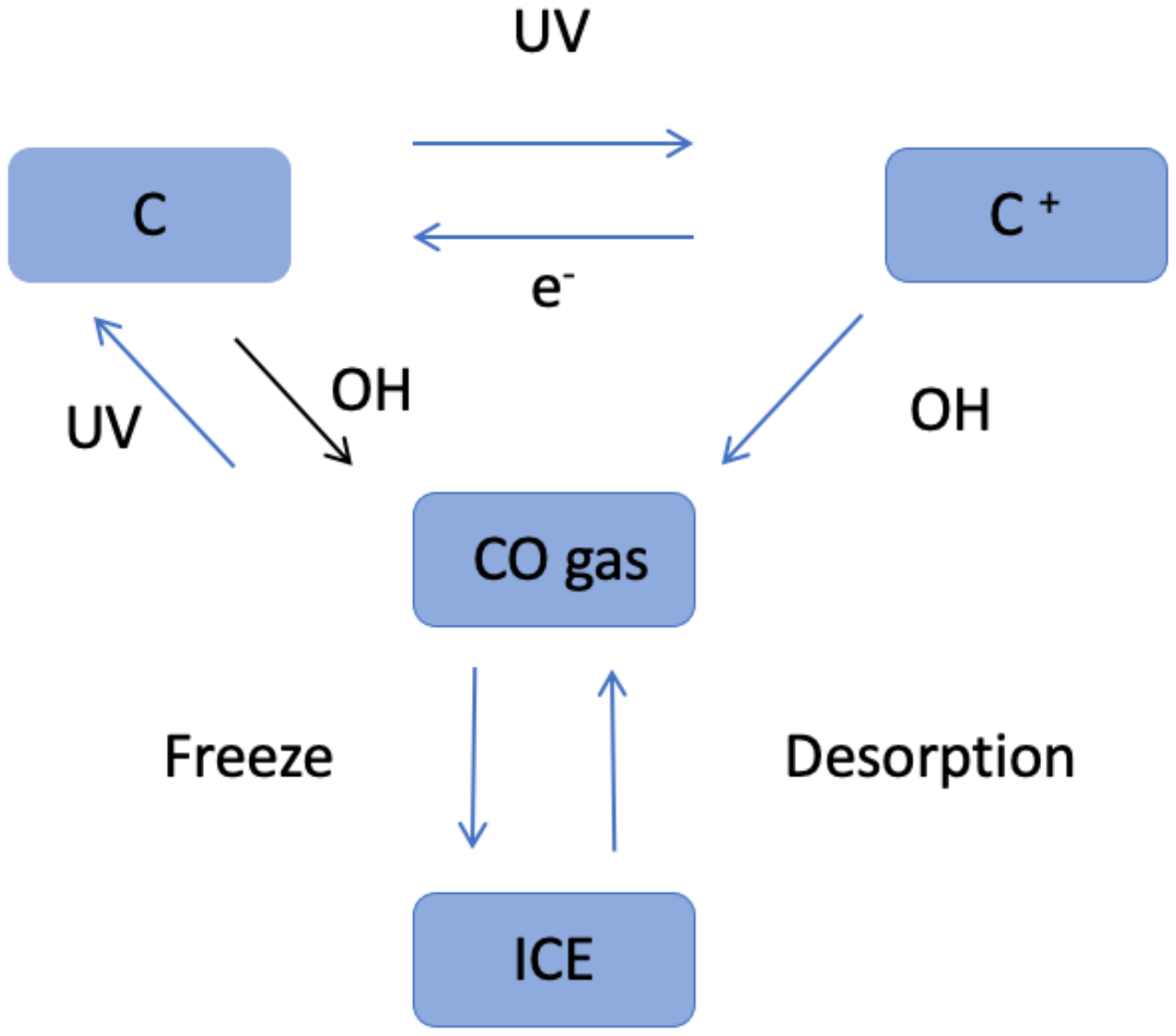}
\caption{Simplified carbon chemistry network.}
\label{fig:COlite}
\end{figure}

KC08 suggested that the equilibrium abundances of CO, C, and C$^+$, and CO ice 
can be approximated by a simultaneous solution of the equilibrium rate equations, and 
that this approximation is adequate for quiescent molecular clouds such as starless cores. 
KRC14 suggested a similar procedure for the abundances of O, OH, H$_2$O and H$_2$O 
ice as discussed in detail in Appendix B of that paper. From the discussion above 
and in \S\ref{subsec:COchem}, we can now define a revised reduced chemical network 
(figure~\ref{fig:COlite}) for carbon and CO. This allows us to simultaneously solve the 
rate equations above for the equilibrium abundances of C, C$^+$, CO, and CO ice 
independently of STARCHEM. To do so, we assume that the abundance of OH is given 
by the reduced network chemistry for oxygen species, and the abundance of free 
electrons is equal to the abundance of C$^+$. The latter results in a 
quadratic equation for C$^+$. 
The carbon-chemistry reaction scheme and recent values for the rate 
coefficients \citep[taken from {\sc UDFA22};][]{udfa22} are:
\begin{align*}
{\rm C + h\nu \to C^+ + e^-} & & 
{\rm 3.5\times10^{-10} e^{-3.76A_v}~s^{-1}}\\
{\rm C^+ + e^- \to C + h\nu} & & {\rm 1.23\times10^{-11}T^{-0.29} 
e^{17.6/T}~cm^3s^{-1}} \\
{\rm C + OH \to CO + H} & & {\rm 1.0\times10^{-10}~cm^3s^{-1}} \\
{\rm C^+ + OH \to CO^+\dots\to CO} & & {\rm 1.33\times10^{-8}T^{-0.5}~cm^3s^{-1}} \\
{\rm CO + h\nu \to C + O} & & {\rm 2.4\times10^{-10} e^{-3.88A_v}~s^{-1}} \\
{\rm CO \to CO_{(ice)} } & & {\rm freeze-out} \\
{\rm CO_{(ice)} \to CO } & & {\rm desorption}
\end{align*}
The rates for the freeze-out and desorption reactions are described in 
\S\ref{sec:freeze} and \S\ref{sec:ggrain}, respectively.
The simple description of ice desorption considers the rate to be proportional to the 
total solid-state abundance of a species and omits the detailed calculations of the 
several grain surface processes in STARCHEM, including the desorption 
due to the energy released by the formation of molecules on grain surfaces. 
Owing to the quadratic equation for C$^+$, we solve the set of simultaneous rate
equations by elementary algebra rather than matrix inversion as was used in KRC14.

\section{The STARCHEM model}
\label{sec:starchem}

Although {\sc STARCHEM} has been
applied in previously published studies \citep[e.g.][]{RW21,Raw22} it was originally
formulated for the study of pre-stellar and protostellar cores, so
we give a fuller description of the model here. 
The model was designed to allow fully flexible descriptions of the 
chemistry and dynamics and the main features of the generic code are as follows:
\begin{enumerate}
\item The chemistry and physical parameters are co-integrated using the 
{\sc DLSODE} integration package, along Lagrangian streamlines, for a
number of test points in the chosen geometry. The model has 
been constructed so as to be easily adaptable to any user-defined 
dynamics.
The physical parameters (such as the density, gas and dust temperatures, 
visual extinction and velocity field) can either be specified analytically,
or in look-up tables, with the continuous values 
determined though smoothed interpolation. 
\item The chemistry is defined by flexible descriptions of 
gas-phase reactions, 
gas-grain interactions (in the surface layers of the dust ices), 
and solid-state (sub-surface) processes. 
In particular, the chemical datasets (species and reaction rate coefficients) 
are dynamically switchable - allowing the sensitivity to the adopted 
chemistry to be tested.
\item Where appropriate (i.e. at relatively low values of extinction), 
the code incorporates a model of the photon dominated region (PDR) to 
describe the photochemistry.
\item Freeze-out, plus a full suite of desorption processes are included
(thermal, direct and CR-induced photodesorption, CR heating - spot and whole grain,
heating by H$_2$ formation and species-specific enthalpy of formation etc.), 
together with a simplified surface chemistry.
\item The chemical composition of the ices that form on the dust grains
are determined with time- and position-dependence on a 
layer-by-layer basis, in a so-called `three-phase' model, which differentiates
the chemically active surface layers and the chemically inert
sub-surface layers.
\end{enumerate}

Details of these various features are given below, as well as the specific 
application to our model of L1544.

\subsection{The gas-phase chemistry}
\label{sec:gasphase}

One of the features of {\sc STARCHEM}, and partly to allay any concerns that the
results are artefacts of errors and inconsistencies in reaction rate data, is its
flexibility to be able to dynamically switch between different chemistries. 
We consider three databases:
(a) {\sc UDFA 2012 }\citep{udfa12},
(b) {\sc UDFA 2006 }\citep{W07} and
(c) {\sc KIDA 2012 }\citep{Wak12}.
As our basis set, we use the first of these. 
In this study, the reaction set consists of some 1250 chemical reactions 
between 98 species, involving the elements H, He, C, N, O, S and a representative
low ionization potential metal, Na. Molecules are limited to species composed 
of less than 6 atoms.
The chemistry includes ion-molecule, neutral-neutral, cosmic ray ionizations and 
photochemical reactions as applicable to molecular clouds. Secondary 
(cosmic ray induced) photodissociation is also included.

In this study we made a few alterations to the reaction data.
e.g. the rates given in {\sc UDFA Rate12} for the reaction 
\[ {\rm H + OH \to H_2O} \]
are probably anomalous (T.J. Millar, private communication):
\[ {\rm k = 5.28\times 10^{-18} (T/300)^{-5.22}\exp (-90/T)}. \]
If this were correct, then due to the strong 
temperature-dependence, this reaction would be a dominant contributor to 
the H$_2$O formation rate (in the low temperature, gas-phase limit).
Instead, we adopt the values given in the Ohio State University (OSU) database:
\[ {\rm k = 4.0\times 10^{-18} (T/300)^{-2.0}} \]

\subsection{The Photon Dominated Region}
\label{sec:pdr}

Photochemistry clearly plays an important role in the chemical evolution
of starless cores, both through gas-phase photolysis and the desorption
mechanisms described below.
In the outer parts of the core ($r\gtappeq 0.07$pc) the mutual shielding of the 
H$_2$, C and CO photodissociation and photoionization continua, coupled to the
dust absorption are very important.
The model includes a simulation for the photon-transmission probability
in a photon dominated region (PDR). This is particularly important for 
the determination of the rates for H$_2$ photodissociation, C photoionization 
and CO photodissociation, and hence the H/H$_2$ and C$^+$/C/CO transitions.
This is a simplified approach that is based on accurate analytical fits 
(as functions of extinction and radiation field strength) to the data 
presented in Figures 8, 10 and 13 of \citet{Retal07}.


The code also includes allowances for the mutual shielding effects.
Specifically, the shielding by the CO photodissociation and C photoionization 
lines/continuum has significant implications for the photochemistry of 
other species; most notably N$_2$ and CN \citep{RR12}. 

\subsection{Freeze-out}
\label{sec:freeze}

Following \citet{RHMW92}, the rate of change of the gas-phase fractional abundance 
of a species, $i$, due to freeze-out onto the surface of grains is given by:
\begin{equation}
\dot{X}_{\rm i} = -\left( \frac{k_{\rm B}}{2\pi m_{\rm H}}\right)^{\frac{1}{2}} 
\sigma_{\rm H}S_{\rm i}C\left( \frac{T}{m_{\rm i}}\right)^{1/2}n_{\rm H}X_{\rm i}
\end{equation}
where: 
$\sigma_{\rm H}$ is the mean surface area of dust grains per hydrogen nucleon
and is the grain population-averaged value of 
$ 4\pi \langle a^2\rangle .d_{\rm g}$.
Here $d_{\rm g}$ is the dust to gas ratio (by number, relative to $n_{\rm H}$) and
the factor of 4 in this expression implicitly assumes that the grains are
spherical.
$n_{\rm H}$ is the hydrogen nucleon density (cm$^{-3}$),
$S_{\rm i}$ is the sticking coefficient,
$T$ is the gas kinetic temperature,
$m_{\rm i}$ is the molecular mass of species $i$ in $amu$,
and C is a factor which takes into account electrostatic effects for ions 
(assuming an average grain charge of -1); C=1 for neutral species, whilst for singly 
charged positive ions $C = 1+ (16.71/aT)$.

Following laboratory measurements and quantum mechanical calculations we assume 
that $S_{\rm i}\sim$1.0 for all species except that the possibility for the
differentiation of ions and neutrals is included.
It is also assumed that electrons impinge upon and stick to grains at a rate 
that is just sufficient to neutralise all accreted ionic species.

\subsection{Desorption}
\label{sec:ggrain}

Comprehensive descriptions of the various continuous desorption mechanisms 
are included in {\sc STARCHEM}. These are:
(i) thermal desorption,
(ii) direct photodesorption by the (attenuated) interstellar radiation field,
(iii) photodesorption driven by radiation field generated by the cosmic-ray
excitation of excited states of H$_2$,
(iv) desorption due to cosmic ray heating of whole grains or ice mantle hot spots,
(v) non-selective chemical desorption driven by the enthalpy of H$_2$ 
formation on grains, and
(vi) species-specific chemical desorption driven by the enthalpy of their 
formation (e.g. for H$_2$O, NH$_3$ etc.).

Most desorption processes involve sublimation from the 
surface layers of the ices that depends on the
compositional structure of the ice.
An ice mantle that is predominantly composed of one molecular species 
(e.g. H$_2$O) may be covered by ice layers of a different composition (e.g. CO) 
that effectively protects the lower levels of the ice against desorption. 
We therefore allow for the detailed time-dependent
variations of the composition in the individual ice layers, 
and adopt the so-called `three-phase' model of the chemistry; 
differentiating between the gas-phase, the `active' surface ice layers and the 
inert ice core.
This implementation is described in more detail in subsection~\ref{sec:surf}.

\subsubsection{Thermal desorption}
\label{sec:thermal}

The rate of thermal sublimation (per molecule) for a zeroth order process 
is given by:
\[ k = \nu_0 e^{-E_{\rm bind}/k_{\rm B}T_{\rm dust}} {\rm ~s}^{-1} \]
where $E_{\rm bind}$ is the binding energy of the adsorbate, 
$T_{\rm dust}$ is the dust temperature and $\nu_0$ is the vibration 
frequency of the adsorbed molecule given by 
\[ {\rm \nu_0 = \sqrt{2N_sE_{bind}/\pi^2m_i} } \]
where $N_{\rm s}$ is the number of binding sites per unit area, and $m_{\rm i}$ 
is the mass of the adsorbed particle. Typically, $\nu_0 = 10^{12}-10^{13}$s$^{-1}$.
This implies an evaporation rate  of 
\[ f_{\rm i}N_{\rm s}\sigma_{\rm H} n_{\rm H}\nu_0 e^{-E_{bind}/kT_{dust} 
\mbox{~~cm$^{-3}$s$^{-1}$}, } \] 
where $f_{\rm i}$ is the fractional surface coverage of species $i$ and 
$\sigma_{\rm H}$ is the grain surface area per hydrogen nucleon, as described above.

The exponential dependence on the dust temperature implies a critical temperature,
primarly defined by the binding energy of the adsorbate. This is rather ill-defined 
in ices of mixed composition although, for the dust temperature profiles given in 
Figure~\ref{fig:physics}, thermal desorption is not expected to be important
except for the most volatile species.

\subsubsection{Photodesorption}
\label{sec:photodes}

The photodesorption of molecules from ices proceeds by
absorption of a far-UV photon (which typically has an energy greater than the
bond energy of the absorbing molecule, and very much greater than the surface 
binding energy of the molecule). This results either in the direct
desorption of the molecule, or else its photodissociation 
{\em in situ}. The dissociation may be followed by the desorption of one or 
more of the dissociation products. Alternatively, the molecule may recombine 
and be chemically desorbed as a result of the energy released in the bond 
formation. 

For H$_2$O the process is driven by solid-state photodissociation. 
This has been studied, both experimentally and theoretically, in some 
detail \citep[e.g. see review in][]{DHN13} 
and in our model we include two possible photodesorption channels:
\[ {\rm H_2O(s) + h\nu \to H_2O }\]
\[ {\rm H_2O(s) + h\nu \to OH + H }\]
although the channel
\[ {\rm H_2O(s) + h\nu \to O + H_2 }\]
may also be possible with lower efficiency.
The laboratory \citep[e.g.][]{OL09,Cruz18}, 
theoretical \citep[e.g.][]{And06} 
and observational 
evidence \citep[based on the observed OH:H$_2$O ratio in translucent clouds, e.g.][] 
{Holl09} all suggest that the yield for dissociative photodesorption is approximately
twice that for non-dissociative desorption, and we split the total yield 
for photodesorption between these processes accordingly.
%
%
Similar branchings are included for other important ice mantle components, such as 
methane:
\[ {\rm CH_4(s.) + h\nu \to CH_2 + H_2 }\] 
\[ {\rm CH_4(s.) + h\nu \to CH_4 }\] 

The UV photons that cause the desorption can only penetrate 
the upper (maybe 3 or 4) monolayers of ices - with quite limited efficiency.
In general, the desorption flux (cm$^{-2}$) due to photodesorption
is given by:
\[ F = Y_{\rm pd,i}F_{\rm UV}f_{\rm i} \]
where $Y_{\rm pd,i}$ is the photodesorption yield for species $i$, 
$f_{\rm i}$ is the fraction of binding sites in the surface layer(s) that are 
occupied by species $i$ (as determined from the model of the compositional 
structure of the ice layers) and $F_{\rm UV}$ is the photon flux, given by:
\[ F_{UV} = G_0F_0e^{-\beta A_{\rm v}} + F_{cr}\left(\frac{\zeta}{\zeta_0}\right), \]
where $\zeta_0=1.3\xten{-17}$s$^{-1}$ is the standard value for the 
cosmic ray ionization rate.
In this expression the first term accounts for direct photodesorption; 
$F_0\sim 10^8$photons cm$^{-2}$ is the unshielded interstellar UV flux, with a
scaling factor $G_0$ and the exponential accounts for shielding by dust. 
$F_{\rm cr}$ is the cosmic ray induced photon flux \citep[from][]{CPA92} and 
accounts for cosmic-ray induced photodesorption. 

There are therefore three major uncertainties in quantifying the 
photodesorption rates:
(i) the (total) yield for the process $Y_{\rm pd,i}$, and its 
dependence on the substrate where the molecule is located,
(ii) the products and branching ratios of the desorption process (if 
dissociative), and
(iii) the $A_{\rm v}$-dependence of the rate ($\beta$).

In the case of H$_2$O all three of these have been well studied.
$Y_{\rm pd}$ has been measured in the laboratory to be 
$\sim 10^{-4}-10^{-3}$ for pure water ices on a flat gold substrate. 
It seems unlikely that the same results would apply to 
impure, mixed ices on amorphous and morphologically complex grain surfaces.
%
%
%
The photodesorption yields for other species, and particularly for ices of 
mixed composition, are poorly understood.
However, it is unlikely that the photodesorption of 
H$_2$O molecules would occur whilst leaving other (perhaps more volatile) species 
bound to the ices, particularly in ices that are predominantly composed of H$_2$O.
So, we make the simplifying assumption that the same photodesorption yield 
applies to {\em all} ice species, although we consider that yield to be a
free parameter. 

The values of $\beta$ are also highly uncertain and depend on the 
(a) the photodesorption efficiency, and (b) the dust opacity, which are both
wavelength dependent.
In the absence of information, we crudely assume (in common with many other 
studies) that the same value of $\beta=1.8$ for H$_2$O applies to all other 
species.
We recognise that this is not well justified and could result
in significant errors in the gas-phase abundances determined in regions 
of moderate extinction ($A_{\rm v}\sim 1-16$). 

\subsubsection{Desorption driven by cosmic-ray heating}
\label{sec:crdes}

The interaction between cosmic rays and ice-coated grains is
complex and ill-defined. Cosmic ray strikes may lead to desorption 
in a number of ways including direct sputtering, heat-spike 
pressure-pulse desorption and via grain heating.
At one extreme, high energy cosmic rays may be capable of 
totally disrupting (small) grains, whilst at the other they 
may simply act to heat the grains.
In this study we adopt the commonly used grain heating 
model - in which cosmic ray impacts, that typically occurring once every 
Myr, result in grain heating and the rapid desorption of material from 
the surface layers, accompanied by efficient evaporative cooling. 
This is a sporadic thermal desorption process
that may result either from whole grain heating or, in some cases, 
spot heating at the entry/exit points of the cosmic ray on the 
grain surface.
The efficiency of the process depends on the binding energies of the 
molecules and the spectrum of the cosmic rays, together with the duty 
cycle of grain heating and cooling.

Recent studies have questioned the various assumptions that go into
these models. These indicate that the previously adopted values of the 
effective desorption rates may have been significantly underestimated
\citep[e.g.][]{Sip19,KK20,SSC21,Raw22} and that the sporadic nature of the 
process cannot easily be reduced to continuous desorption rates 
\citep{Raw22}.
For the ease of comparison, and to test our stated
objectives, we have continued to use the updated, continuous desorption 
rates determined for whole grain heating. We add a note of 
caution concerning their validity.

With this assumption, the rates can be expressed \citep[e.g.][]{HH93} as:
\[ \dot{n}=R_{\rm cr}n_{\rm i} \]
where $R_{\rm cr}$ is the desorption rate per molecule and $n_{\rm i}$ is
the {\em volume} abundance of solid-state species $i$.

There are several variations in the calculated values for the 
desorption rates. For example,
by considering more localised ice mantle `hot spot' heating, \citet{BJ04}
calculated a rate for H$_2$O desorption that is approximately 30 times
larger than the value given in \citet{HH93}. 
We use these `hot spot' values for H$_2$O, and other species, in our model. 

\subsubsection{Desorption driven by H$_2$ formation}
\label{sec:h2des}

The formation of H$_2$ molecules on the surface of grains yields an energy
equal to the bond energy of H$_2$. A significant fraction of this energy may be 
deposited locally in the substrate and conducted to the rest of the grain.
This energy may be sufficient to desorb other nearby molecules, non-selectively 
\citep{WRW94,Pant21}. 
As H$_2$ formation occurs on the surface of grains the
desorption process is limited to molecules in the surface layers.
In this case, the rate of desorption (cm$^{-3}$s$^{-1}$) of a species $i$ is
given by
\[ \dot{n}_{\rm i} = Y_{\rm H_2,i}.\dot{n}_{\rm H_2}.f_{\rm i} \]
where $Y_{\rm H_2,i}$ is the desorption yield for species $i$, 
$f_{\rm i}$ is the surface coverage of species $i$, as defined above, and 
$\dot{n}_{\rm H_2}$ is the rate of (surface) 
H$_2$ formation, which we determine from the hydrogen atom
impingement rate on grains.

\subsubsection{Desorption driven by the enthalpy of formation}
\label{sec:chemdes}

We also include the selective desorption of individual molecules, 
driven by the enthalpy of their formation on the surfaces of 
bare or ice-coated grains. This is important for several species, such as H$_2$CO
and - most especially - OH and H$_2$O. To quantify this, we use the results of
\citet{Min16} who determined the chemical desorption efficiencies to
be $\sim 50$\%, 
amorphous water ice substrates respectively, and $\sim 50$\%, $\sim$80\% and 
$\sim$ 30\% for the OH+H$\to$H$_2$O reaction on oxidised graphite, amorphous 
silicates and amorphous water ice, respectively.
This dependence of the desorption efficiencies on the nature of the surface
is included in the model. 
Additional complexities, such as the likelihood of higher desorption efficiencies
from CO-rich ices \citep{Vas17} have not yet been included in the model.

\subsection{Surface chemistry}
\label{sec:surf}

In the current version of {\sc STARCHEM} we do not consider the thermodynamics 
and kinetics of surface chemical processes, which are dominated by the thermal 
hopping of hydrogen and other, light, reactive species. Instead we apply a limited
and very simple realisation of the surface chemistry, and consider the fractional 
conversion of species, subject to empirical constraints.

Thus, all neutral molecules and radicals are allowed to stick to grains and,
where appropriate, hydrogenate to saturation 
(e.g. C, C$^+$, CH, CH$_2$, CH$_3 \to$ CH$_4$ and N, N$^+$, NH, NH$_2 
\to$ NH$_3$ etc.).
Some atomic ions (He$^+$, Na$^+$ and S$^+$) are allowed to recombine 
and return to the gas phase, whilst all molecular ions are assumed to 
dissociatively recombine, with the products returned to the gas-phase. 
H$_2$ formation is described using the same formalism as for other species.
For the temperature-dependence of the combined sticking and reaction probability 
for the conversion of H to H$_2$ we use the simple formula of \citet{BZ91}.

For the surface chemistry that results in the 
formation of H$_2$O and CO$_2$ we note the following:
(a) H$_2$O and CO$_2$ are not very reactive and so can be treated as being 
chemically inert on dust grains,
(b) \citet{Pont06,GP11} suggest that CO$_2$ 
is formed on a CO ice surface via the surface reaction of O or OH with CO, 
and  
(c) empirically, for Taurus cores, H$_2$O is the dominant (observed) ice,
with the CO and CO$_2$ abundances relative to H$_2$O being $\sim$10-40\% 
and $\sim$20\% respectively \citep{Chiar11}.

For the oxygen chemistry, the freeze-out and surface reactions of 
significance are (representing solid-state species by `(s)'):
\begin{equation}
 {\rm O/O^+ + grain \to O(s) }
\label{eqn:Ofreeze}
\end{equation}
\begin{equation}
 {\rm OH + grain \to OH(s) }
\label{eqn:OHfreeze}
\end{equation}
\begin{equation}
 {\rm O, O^+, OH + CO(s) \to CO_2(s) }
\label{eqn:CO2form}
\end{equation}
\begin{equation}
 {\rm O(s) + H \to OH(s)/OH }
\label{eqn:OHform}
\end{equation}
\begin{equation}
 {\rm  OH(s) + H \to H_2O(s)/H_2O }
\label{eqn:H2Oform}
\end{equation}
\begin{equation}
 {\rm O_2 + grain \to O_2(s) }
\label{eqn:O2freeze}
\end{equation}
\begin{equation}
 {\rm O_2(s) + H ~....\to H_2O(s)/H_2O }
\label{eqn:O2hyd}
\end{equation}
The chemical endpoint for the oxygen surface chemistry is thus dominated 
by water formation.

For the very low dust temperatures within starless cores, CO is much less 
mobile than hydrogen atoms on grains surfaces \citep{GP11}. So, to allow 
for the formation of CO$_2$(s) (as opposed to hydrogenation products)
we assume that CO$_2$(s) is formed by the non-diffusive prompt 
(Eley-Rideal) reaction of gas-phase O, O$^+$ or OH with bound CO molecules (reaction~\ref{eqn:CO2form}).
The impact probability is thus proportional to the surface coverage of CO, 
$f_{\rm CO}$, and the reaction efficiency is $f_{\rm CO}.F_{\rm CO_2}$, where 
$F_{\rm CO_2}$ is the reaction probability per impact.
In typical simple models of gas-grain interactions $F_{\rm CO_2}=0.1$ is
typically adopted to account for the observed CO$_2$:H$_2$O abundance ratio. 
For the O, O$^+$ or OH that does not react with CO, and providing the timescale 
for the surface residence of adsorbed oxygen atoms is greater than the hydrogen 
atom migration timescale over the surface, we assume that the hydrogenation reactions
(\ref{eqn:OHform} \& \ref{eqn:H2Oform}) are efficient, so that a fraction
($F_{\rm reac}\sim 1$) is hydrogenated to form H$_2$O.

A proportion of the surface-formed OH and H$_2$O may be desorbed due to the 
enthalpy of formation ($D_{\rm OH}$ and $D_{\rm H_2O}$, respectively). 
The efficiencies of these processes are discussed and quantified
in \citet{Min16} and \citet{Vas17}, as discussed above.
Following that work, we can include formation enthalpy-driven 
desorption with $D_{\rm OH}=0.5,0.25$ and $D_{\rm H_2O}=0.8,0.3$ for formation on
bare, refractory, grains and amorphous H$_2$O ice, respectively.
For O$_2$ we assume that a fraction ($F_{\rm O_2}$, typically $\sim$1) is
ultimately hydrogenated to H$_2$O (reaction~\ref{eqn:O2hyd}) via several possible 
chemical pathways, that we do not discuss and, again, a proportion ($D_{\rm H_2O}$) 
of this H$_2$O is desorbed.
The remaining O, O$^+$, OH and O$_2$ not undergoing these reactions is simply assumed 
to freeze out (reactions~\ref{eqn:Ofreeze}, \ref{eqn:OHfreeze} \& \ref{eqn:O2freeze}).
The binding energies ($E{\rm _B}$/k) for O, OH, O$_2$ and H$_2$O are
taken to be 800K, 1300K, 1210K and 5770K, respectively,
although some authors claim that the binding energy of oxygen 
atoms may be as high as 1660 or 1850K on amorphous ice and bare refractory 
grains, respectively \citep{He15}.
Our approach allows for the various poorly-defined competing processes of 
freeze-out, desorption, hydrogenation and more complex reactions. 

For the carbon chemistry, other than CO, the dominant species in 
the freeze-out/hydrogenation reactions will be C/C$^+$. The intermediate 
species CH has a very low binding energy ($\sim$650K) and so it is likely 
that enthalpy-of-formation driven desorption may be important:
\[ {\rm C,C^+ + grain \to CH }\]
We treat the efficiency of this process as a free parameter ($f_{\rm CH}$).

We should also include the hydrogenation of CO, leading to the formation of 
methanol (CH$_3$OH). This appears to be a very efficient mechanism on the 
surface of dust grains at low temperatures \citep[see eg.][]{Vas17,Ried23}.
We do not, however, follow the chemistry of CH$_3$OH in this study.

\subsection{Ice composition and layering}
\label{sec:layers}

In {\sc STARCHEM} we pay careful attention to the composition of the ices 
and trace this composition on a microscopic, layer-by-layer basis.
We recognise that both the freeze-out of gas-phase species and most 
desorption processes result in the addition, or removal, of the top 
layer(s) of the ice leaving the sub-surface layers unaffected. 

This implies that, at any given time, the nature of the gas-grain interactions
is determined by the structure and composition of only the surface layers.
This is the only chemically active part of the ice, within which surface
migration and reaction can occur; the sub-surface layers are
treated as being chemically inert and, in most situations, chemically 
inaccessible. We do not allow for any diffusion between the layers.
This is similar to the so-called `three phase' approach adopted by in other
studies \citep[e.g.][]{TCK12}.
In regions where there is progressive freeze-out the sub-surface layers will
be a `fossil record' of the prior chemical/dynamical evolution of
the cloud. Unless the ices are disrupted or sublimated, the subsurface layers
are removed from the active chemistry and the net budget of accessible volatiles.
This implies that, even in those situations where chemical
quasi-equilibrium exists, the observed gas-phase abundances will be dependent on 
the prior evolution of the cloud.
In addition, the composition of the surface layers may be quite different to 
that of the bulk ice. In these circumstances, the effect of the outer layers
`shielding' the inner layers against surface desorption results in significant
imbalances between the 
gas-phase and solid-state abundances of species. This is an idea that has been 
explored in the context of the anomalous O$_2$ abundances seen in comet 67/P 
\citep{RWW19}.
Moreover, the compositional differentiation may also lead to significant errors 
in the estimation of solid-state abundances from observational data.

The determination of the composition of the ice mantles in environments
where ice layers can be added or removed is a non-trivial computational 
exercise but is one that is essential if we want to correctly quantify 
the gas-grain interactions.
The chemically significant parameter at 
any one instant is the number of surface binding sites per grain, which is 
given by
\[ 4\pi\overline{a}^2N_{\rm s}\]
where $N_{\rm s}$ is the surface density of binding sites (cm$^{-2}$)
or, expressed as a `fractional abundance of binding sites' (i.e. the number of
binding sites per hydrogen nucleon):
\[ f_{\rm sites} = \sigma_{\rm H}N_{\rm s}. \]
This parameter is, of course, not fixed and will vary from layer to layer,
as the radius of the grain+ice, and hence the surface area per hydrogen nucleon 
($\sigma_{\rm H}$) will vary as ice layers are deposited or removed.
The fractional coverage of the surface layers by species $i$ is then given by
\[ f_{\rm i} = X_{\rm i,s}/f_{\rm sites} \]
where $X_{\rm i,s}$ represents the fractional abundance of species $i$, apportioned
to the surface layers.

Although this is similar to some other studies \citep[e.g.][]{TCK12}
we do not include some of the additional complexities of those models
(e.g. the inclusion of grain porosity/lattice defects etc., as appropriate for 
`fluffy aggregate' grain models) as there are a 
number of assumptions and simplifications in (all of) the ice layering models 
that are probably more significant.
These include:
(i) the assumption that dust grains are spherical,
(ii) the assumption that ices form in neat, sequential, concentric, layers. 
It is possible that the ice layers may not stack up neatly and there may be
some parts of a dust grain that have many more layers of ice than other parts,
(iii) whilst the segregation of layers implies that each layer
will have its own chemical composition, there is no allowance for any compositional
variations {\em within} each ice layer. We might not expect polar and 
apolar components to be fully mixed and there may be `patches' of ices with different
composition and binding properties as a result, affecting both the freeze-out and 
desorption properties of the surface layers, and
(iv) the assumption that the ice layer structure remains unaltered, other than as a
result of freeze-out or desorption. This could break down, for example, in the case
of energetic cosmic ray heating which acts to melt/partly sublimate the ices and 
hence significantly disrupt the integrity of the ice layers. 


We treat each ice layer as having a (defined) constant thickness equal to the 
mean bond length in the ice. This allows us to calculate 
how the grain radius including the ice mantle (and hence the 
grain surface area) varies with the number of accreted layers. As the freeze-out 
(and surface desorption) rates are proportional to the surface area, we can see how
- in a region where freeze-out dominates over desorption - the rate of freeze-out
can rapidly increase (the `snowball effect').
The accretion of ices may have other effects whose
implications we defer to a future study. These include, for example, the effective
removal of the smallest grains from the grain size distribution 
\citep[e.g.][]{Sil20} and variations in the refractive index and albedo of the grains.

Using the method described above, we can determine the fractions 
of the binding sites in the upper layers that are occupied by the different species, 
with the possible inclusion of additional factors, such as allowances for
variations in the binding energy of key adsorbates as a function of surface 
layer composition.
Ultimately, this allows us to calculate the surface desorption rates and the 
efficiencies of the surface reactions, such as those that result in the formation 
of H$_2$, H$_2$O and CO$_2$.
%

\label{lastpage}

\end{document}